\documentclass{jfm}
\usepackage{graphicx}
\usepackage{epstopdf, epsfig}
\usepackage{newtxmath}
\usepackage{hyperref}

\shorttitle{Stopping vortices from mounted wedge}
\shortauthor{J. C. Kalita}

\title{Simulation of stopping vortices in the flow past a mounted wedge }

\author{Jiten C. Kalita\aff{1}	
	\corresp{\email{jiten@iitg.ac.in}}}

\affiliation{\aff{1}Department of Mathematics, Indian Institute of Technology Guwahati, Guwahati 781039, Assam, India}

\begin{document}

\maketitle

\begin{abstract}
    This work is concerned with the numerical investigation of the dynamics of stopping vortex formation in the uniform flow past a wedge mounted  on a wall for channel Reynolds number $Re_c=1560$. The streamfunction-vorticity ($\psi$-$\omega$) formulation of the transient Navier-Stokes (N-S) equations have been utilized for simulating the flow and has been discretized using a fourth order spatially and second order temporally accurate compact finite difference method on a nonuniform Cartesian grid developed by the author. The results are validated by comparing the simulated results of the early evolution of the flow with the experimental visualization  of a well-known laboratory experiment of \cite{pullin1980} and a grid-independence study. The development of the stopping vortex and its effect on the starting vortex are discussed in details. The stopping flow is analysed in the light of the time interval through which the inlet velocity of the flow is decelerated. The criterion for the development of a clean vortex is provided in terms of the impulse associated with the deceleration. Our study revealed that the strength of the stopping vortex depends upon the rapidity of deceleration. The vorticity distribution along the diameter of the core of the stopping vortex is seen to follow a Gaussian profile.
\end{abstract}
\begin{keywords}

\end{keywords}

\section{Introduction}\label{intro}
 Vortex rings are one of the most fascinating forms found in vortex flows and have attracted the attention of the researchers in the past decades \citep{lim1995, weigand1997evolution,wakelin1997,das2017}. They occur very frequently in nature, the most familiar being cigarette smoke-ring ejected through the mouth of a smoker. They can also be generated in the laboratory by impulsively ejecting fluid through structures resembling orifices into a quiescent fluid (\cite{didden1979formation,auerbach1987experiments,irdmusa1987influence,auerbach1991stirring,lim1995,weigand1997evolution,allen2002interaction,cater2004interaction,das2017}). A brief review of such methods can be found in the work of \citep{das2017}. Most of these methods revealed the generation of a secondary or stopping vortex ring when the flow is suddenly stopped after the formation of the primary or the starting vortex. The characteristics of these vortices were seen to be dependent on the method being employed to generate the starting vortex and the mechanism of stopping the flow. 

 While plethora of experimental visualizations are available in the existing literature on the stopping vortex emanating from orifices, such experiments on flow past a flat plate or wedge normal to the free stream are scantily available in the existing literature. The current study is motivated by the experimental visualization of the starting and stopping flow  past a wedge mounted on a wall by \citet{pullin1980}. Readers may refer to figure 1 \citet{pullin1980} for the documentation of the experimental set-up used in their laboratory experiment.  It contained a channel of height $H=25.5cm$ and width $W=20.3cm$ joined at one end of an open reservoir filled with water. The water level $L_w$ in the reservoir was maintained in such a way that $L_w>H$ in order to make the channel completely full during the experimental runs. A rectangular piston at the other end of the channel performed (readers could also refer to the shaded part in figure 1 of \citep{kalita2023}) the dual duty of driving as well as imparting acceleration to the flow.  The wedge in the form of an isosceles triangular prism of height $h=12.70cm$ was mounted on the upper wall of the channel and formed an angle $\theta=\beta\pi$ at the wedge-apex, called the wedge-angle. The cases undertaken in the current study corresponds to  $\beta=\frac{1}{3}$ so that $\theta=60^{\circ}$ (see figure \ref{fig:stat}).   
 
 \begin{figure}
		\begin{center}
			\includegraphics[width=5.5in]{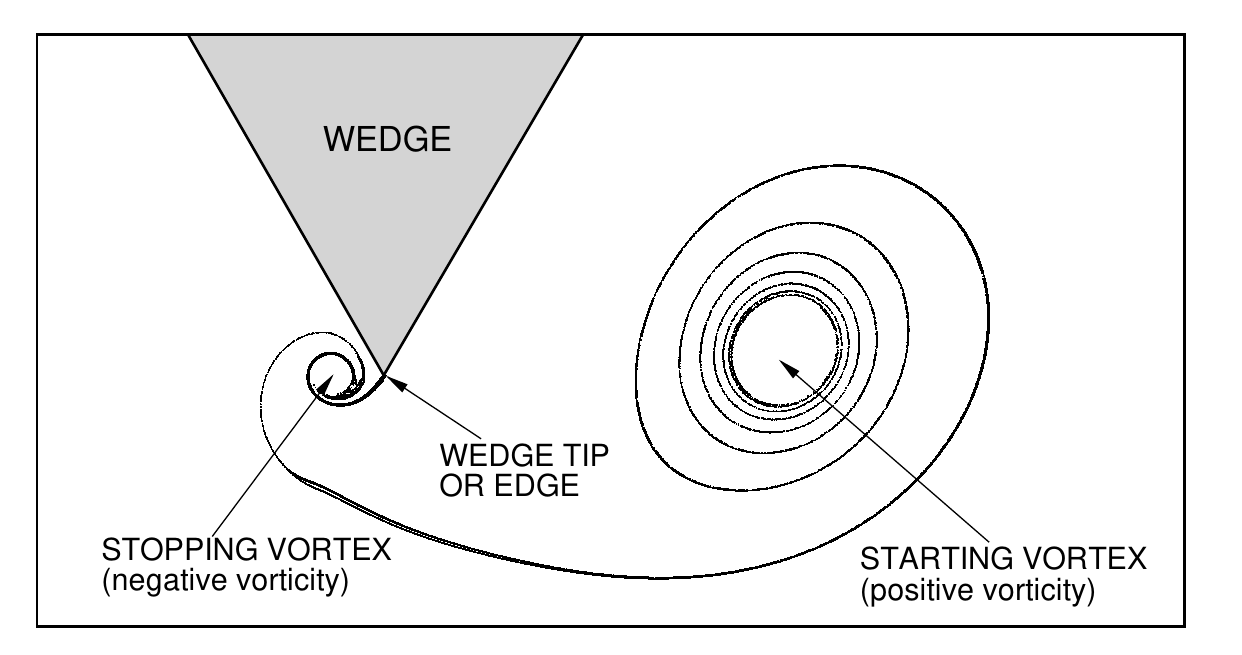}
            %\vspace{-1.5cm}
			\caption{The schematic of the starting and stopping vortices.}
			\label{fig:ss_sch}
		\end{center}
	\end{figure}
 We consider a special case of the accelerated flow considered by them in the laboratory, where the flow reduces to a uniform one with channel Reynolds number $Re_c=1560$ when the pre-chosen constants $A$ and $m$ (would be discussed in the next paragraphs) used by them were assigned values $0$ and $0.63 cm/sec $ respectively.  The schematic of the starting and stopping vortex is shown in figure \ref{fig:ss_sch}. The sharp wedge-edge causes the separation of the flow, paving way for the formation of the starting vortex. When the incoming flow is stopped, a stopping vortex emanates from the same edge with the sign of vorticity opposite to the previously developed starting vortex (figure \ref{fig:ss_sch}).  It is interesting to note that the visualization \citet{pullin1980} mainly focused on and documented the development of a two-dimensional starting flow vortex formed at wedge like sharp edge. Although they presented some results for the stopping flow, they were only in the shape of a few slides. Apart from the time at which the flow was stopped, no more details were available, be it about the mechanism of stopping the flow or the formation of the stopping vortex. The current research is an endeavour towards investigating the effects of stopping the incoming flow and subsequently, studying the flow characteristics inherent to the flow through numerical simulation. In particular, we are interested in the effect of the length of the time interval through which the incoming flow is decelerated to a complete halt. The behaviour of both the starting and stopping vortex is discussed in details. 
   
  \section{The governing equations and their discretization}\label{discr}
	We use the streamfunction-vorticity ($\psi$-$\omega$) formulation (\cite{batchelor2000,ander95,chung}) of the Navier-Stokes equations which governs the flow for incompressible viscous flows. In dimensional form, they can be written as
	\begin{equation}\label{psi_omega}
    \nabla^2 \tilde{\psi}=-\tilde{\omega}, \quad
    \frac{\partial \tilde{\omega} }{\partial \tilde{t}}+\tilde{u}\frac{\partial \tilde{\omega}}{\partial \tilde{x}}+\tilde{v}\frac{\partial \tilde{\omega}}{\partial
    \tilde{y}}=\nu\bigg(\frac{\partial^{2}\tilde{\omega}}{\partial
    \tilde{x}^{2}}+\frac{\partial^{2}\tilde{\omega}}{\partial \tilde{y}^{2}}\bigg).
    \end{equation}
    where the quantities $\tilde{t}$, $\tilde{\psi}$, $\tilde{u}$, $\tilde{v}$  are respectively time, streamfunction, and velocities along the $\tilde{x}$- \& $\tilde{y}$-directions and $\nu$ is the kinematic viscosity. Their units are $\tilde{t}\; (s)$, $\tilde{\psi} \; ({cm}^2 s^{-1})$, $\tilde{u} \; ({cm}s^{-1})$, $\tilde{v} \; ({cm}s^{-1})$ and $\nu \; ({cm}^2 s^{-1})$.
	
	As mentioned above, the flow is driven by the piston which moves with a velocity $\tilde{v}_p=\tilde{u}_0+A\tilde{t}^m$ where $\tilde{u}_0$ (set as zero in the actual experiment) is the 	velocity at the inlet, $A$ and $m$ are pre-chosen constants. Although $\tilde{v}_p$ is responsible for the  acceleration of the viscous fluid, in the current study, the chosen value of $m=0$ facilitates a uniform velocity at the inlet indicating an impulsive start. Equations \eqref{psi_omega} are non-dimensionalized using the channel width $H$ and, the constants $A$ and $m$. Defining the non-dimensional variables
	
	\begin{equation}\label{nondim}
		(x,y)=\frac{(\tilde{x},\tilde{y})}{H},\; (u,v)=\frac{(\tilde{u},\tilde{v})}{A}, \; \psi=\frac{\tilde{\psi}}{AH}\; {\rm and } \; t=\frac{A\tilde{t}}{H},
	\end{equation}
	equation \eqref{psi_omega} in non-dimensional form can be written as 
	\begin{equation}\label{non_sv}
    \nabla^2\psi=-\omega,\;\;\frac{\partial \omega}{\partial t}+u\frac{\partial \omega}{\partial {x}}+v\frac{\partial \omega}{\partial {y}}-\frac{1}{Re_c}\nabla^2\omega =0,
    \end{equation}
    where $\displaystyle Re_c$ is the channel Reynolds number defined as $\displaystyle Re_c=\frac{AH}{\nu}$.

     In order to discretize the $\psi$-$\omega$ form of the N-S equations \eqref{non_sv}, we have used the High Order Compact (HOC) schemes for the 2D convection diffusion (CD) equation on nonuniform grids without transformation developed by the author and his group (\cite{kalita2004,kalita2008}). For the transient CD equations, the temporal accuracy of the scheme is two and for both the transient and the steady-state equations, the spatial accuracy of the schemes lies between three and four. The unsteady convection-diffusion equation in 2D for a flow variable $\phi$ can be written as
\begin{equation}\label{eq:unsteady_general}
\lambda \dfrac{\partial \phi}{\partial t} - \nabla^2 \phi +c(x,y,t)\dfrac{\partial \phi}{\partial x} + d(x,y,t) \dfrac{\partial \phi}{\partial y} = f(x,y,t)
%b\phi_t - (\phi_{xx}+\phi_{yy})+c(x,y,t)\phi_x + d(x,y,t)\phi_y = f(x,y,t)
\end{equation}
where $\lambda>0$ is a constant, $c$ and $d$ are convection coefficients in the $x$- and $y$-directions, respectively, and $f$ is a forcing function. The scheme is given as
\begin{equation}\label{eq:scheme_unsteady}
\sum_{k_1=-1}^1 \sum_{k_2=-1}^1 w_{i+k_1,j+k_2}\phi_{i+k_1,j+k_2}^{(n+1)} = \sum_{k_1=-1}^1 \sum_{k_2=-1}^1 w'_{i+k_1,j+k_2}\phi_{i+k_1,j+k_2}^{(n)} + (0.5F_{ij}^{(n+1)}+0.5F_{ij}^{(n)})\Delta t
\end{equation}
Where $\Delta t$ is the uniform time step and $x_f$, $x_b$, $y_f$, $y_b$ are the nonuniform forward and backward step-lengths in the $x$ and $y$ directions, respectively. Further, $(n)$ and $(n+1)$ stands for the $n$ and $n^{\rm th}$ time levels respectively. The details of scheme development along with the coefficients $w$, $w'$ and the expressions $F_{ij}$ pertaining to the forcing function $f$ in \eqref{eq:unsteady_general} can be found at \cite{kalita2008}. Here, the time discretization follows a Crank-Nicolson strategy, resulting in an unconditionally stable implicit finite difference scheme. Note that letting $\lambda=Re_c$, $c=Re_c u$, $d=Re_c v$ and $f=0$, \eqref{eq:unsteady_general} reduces to the vorticity transport equation of \eqref{non_sv}. Likewise, letting $\lambda=c=d=0$ and $f=\omega$ in  \eqref{eq:unsteady_general} yields the streamfunction-vorticity equation of \eqref{non_sv}, which is discretized by the steady-state version of \eqref{eq:scheme_unsteady}, details of which could be found in \cite{kalita2004}.   
    \begin{figure}
     \hspace{-0.5cm}
       \includegraphics[width=\textwidth]{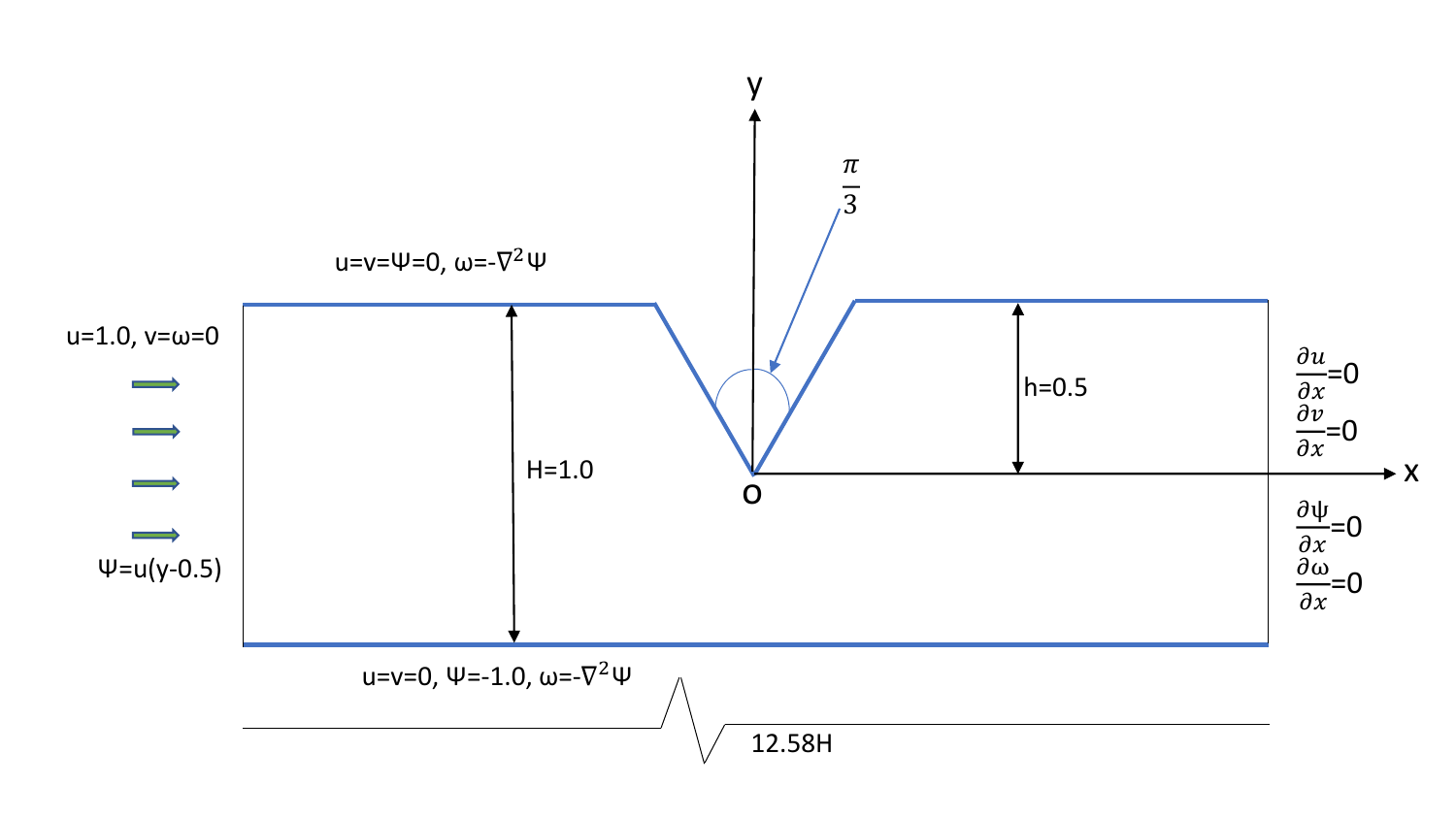}
            %\vspace{-2.5cm}
			\caption{Problem statement of the flow past a mounted wedge in uniform flow.}
			\label{fig:stat}
		\end{figure}
	
	The schematic of the problem under consideration along with boundary conditions used in our computation is shown in figure \ref{fig:stat}.  As shown in the figure, the incoming flow moves from left to right.  Note that use of equation \eqref{nondim} for non-dimensionalization renders the piston velocity $\tilde{v}_p$ used by \cite{pullin1980} to $v_p=1$.  The dimensions of the computational domain was set as $-6.29\leq x \leq 6.29$ and $-0.5\leq y \leq 0.5$ such that the width of the channel is unity and the tip of the wedge is located at the origin.  Following \citet{pullin1980}, the ratio of the channel width and wedge-height is kept at $2:1$ as well. The boundary conditions at the inlet is taken as $u= 1.0,\  v=\omega=0$ and $\psi = u(y-0.5)$. $\psi$ is scaled this way in order to attain a value zero on the top wall, i.e., $\psi_{top}= 0$ and the same streamline corresponding to $\psi=0$ continues its journey by touching the wedge surface. Consequently  boundary conditions for $\psi$ at the bottom wall are $\psi_{bottom}= -1.0$; also $u =0 ,\; v=0$. At the outlet, we have used the zero gradient boundary condition $\dfrac{\partial \psi}{\partial x}= \dfrac{\partial u}{\partial x}=\dfrac{\partial v}{\partial x}=\dfrac{\partial \omega}{\partial x}=0 $. At the top and bottom walls, barring the wedge surface, the vorticity $\omega$ is approximated by one sided approximation of the velocities and the streamfunctions in the vertical directions and central difference in the horizontal direction as in \cite{kalita2008} (see equation (25)). Likewise, for the wedge wall, on the slanted surfaces, suitable one-sided approximation was used in both the directions and on the tip of the wedge, central difference in the horizontal and one-sided approximation in the downward vertical direction were used. 

    We have used a Cartesian nonuniform grid such that in the vicinity of the wedge tip, the grid is extremely clustered. The details of generating such a grid could be found in \cite{kalita2023}, where the grid has been generated in such a way that the wedge boundary passes through the grid points giving rise to a body-fitted coordinate system. 

 \subsubsection{Grid independence and code validation}
\begin{figure}
\begin{center}
	\includegraphics[width=0.4\textwidth]{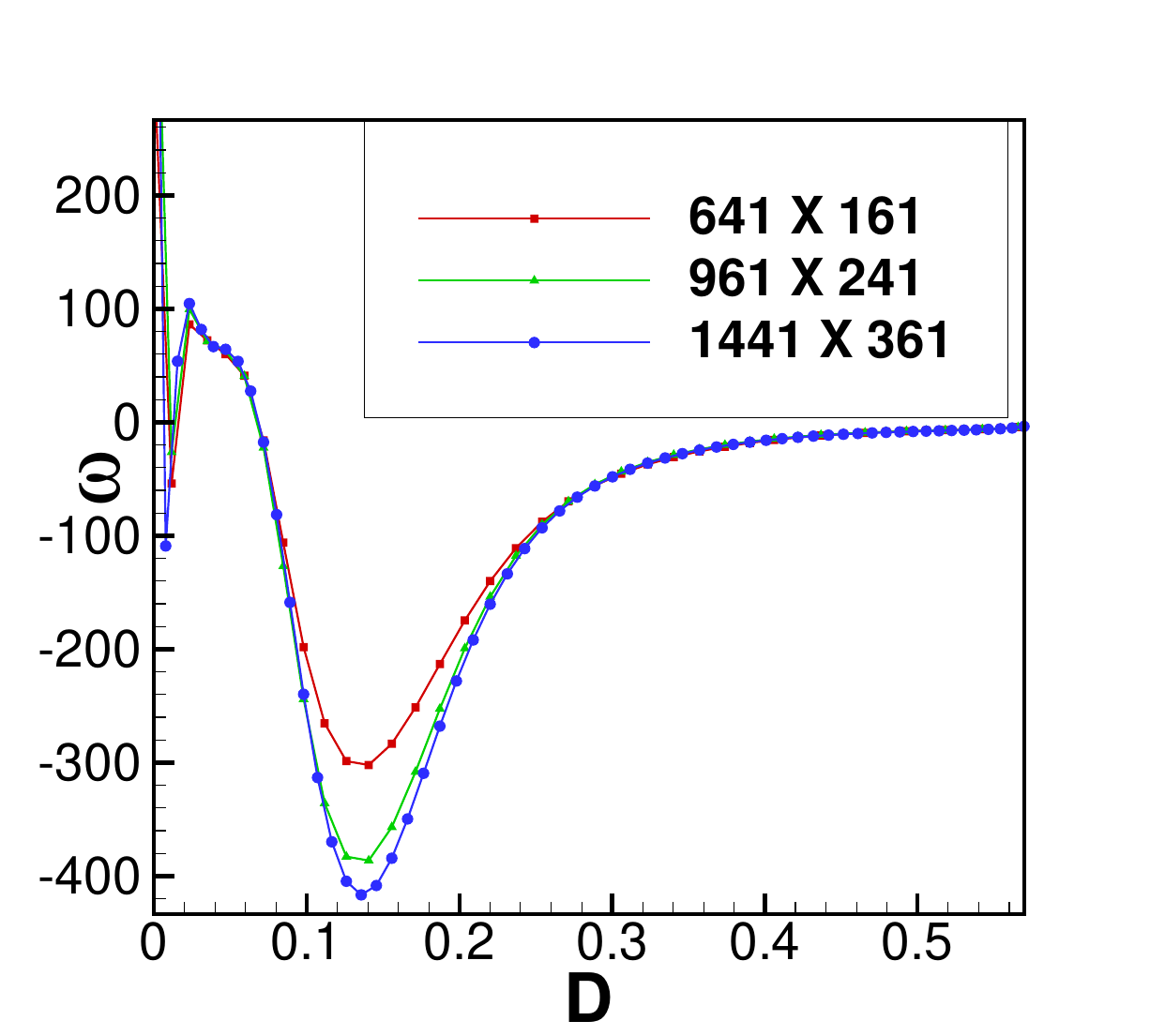}
	{(a)}
	\includegraphics[width=0.4\textwidth]{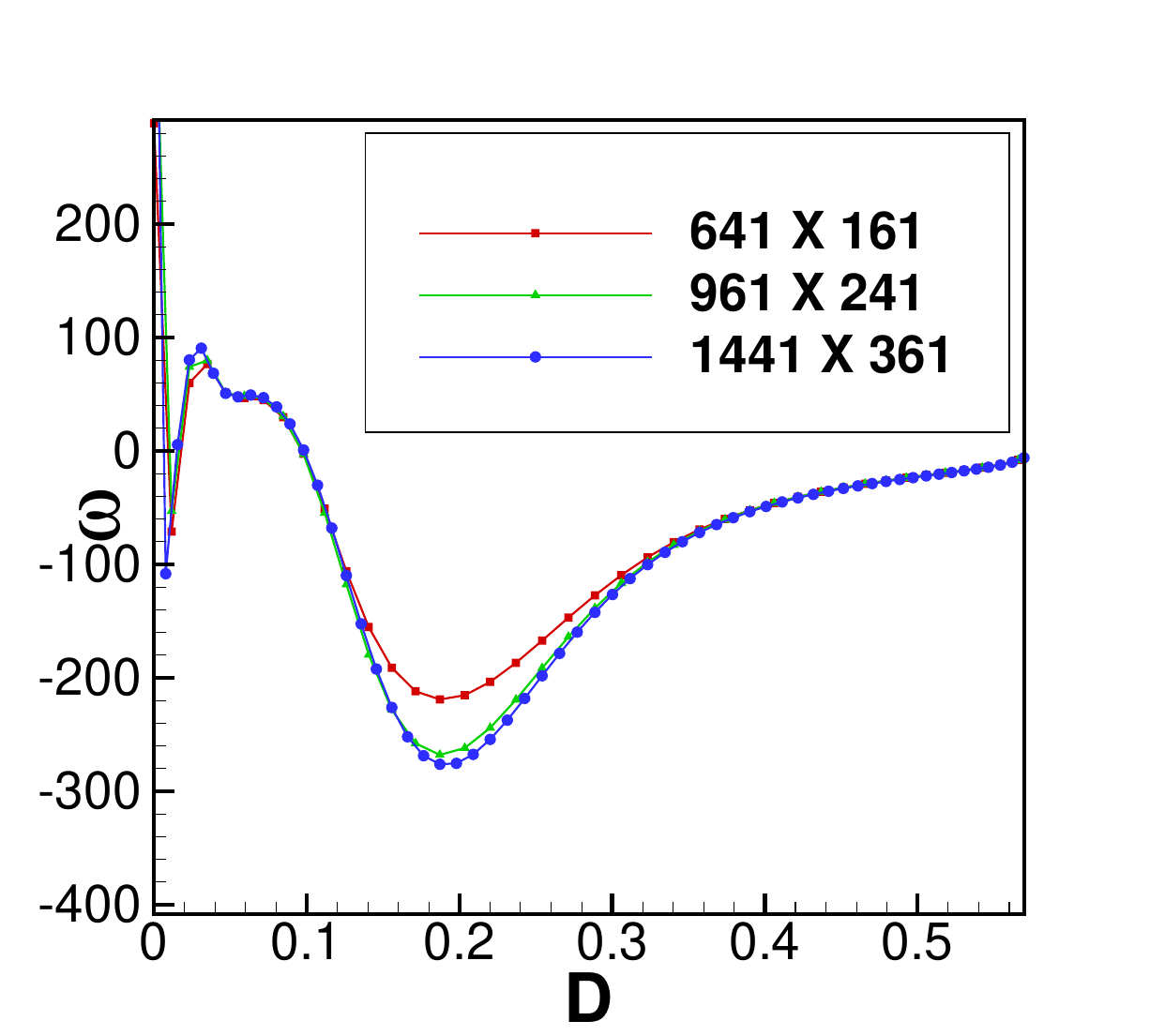}
	{(b)}
	\includegraphics[width=0.4\textwidth]{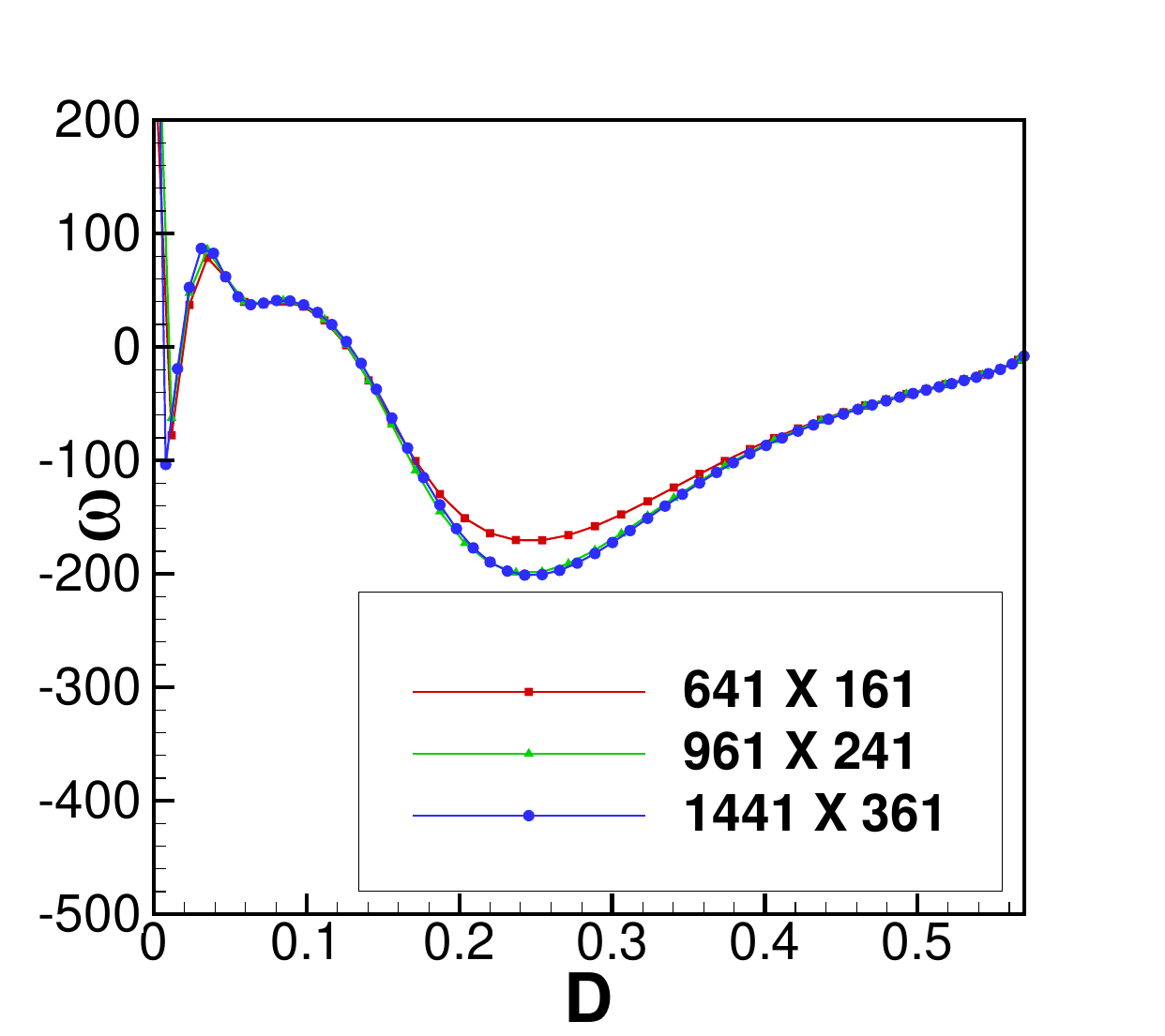}
	{(c)}
	\includegraphics[width=0.4\textwidth]{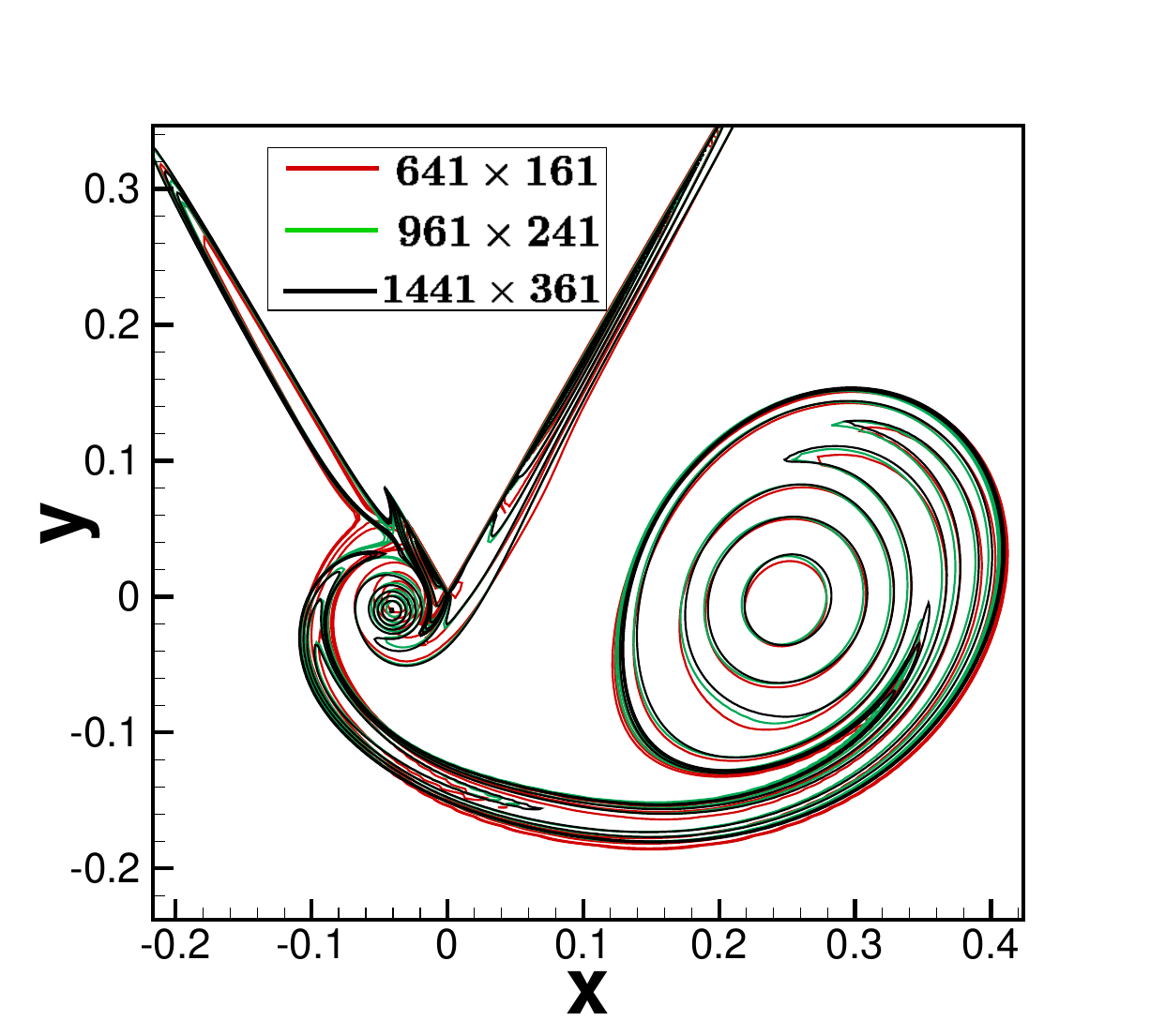}
	{(d)}
	\caption{Grid independence of the computed solutions, on grids of size $641 \times 161$, $961 \times 241$ and $1441 \times 361$: (a)-(c), vorticity distribution along the wedge surface at time $t=0.2$, $0.3$ and $0.4$ respectively, and (d) Vorticity contours depicting the stopping and the starting vortices at time $t=0.52$ corresponding to $t_{dec}=0.02$.}
	\label{grid_ind}
	\end{center}
    \end{figure}

\begin{figure}
\includegraphics[width=5.5in]{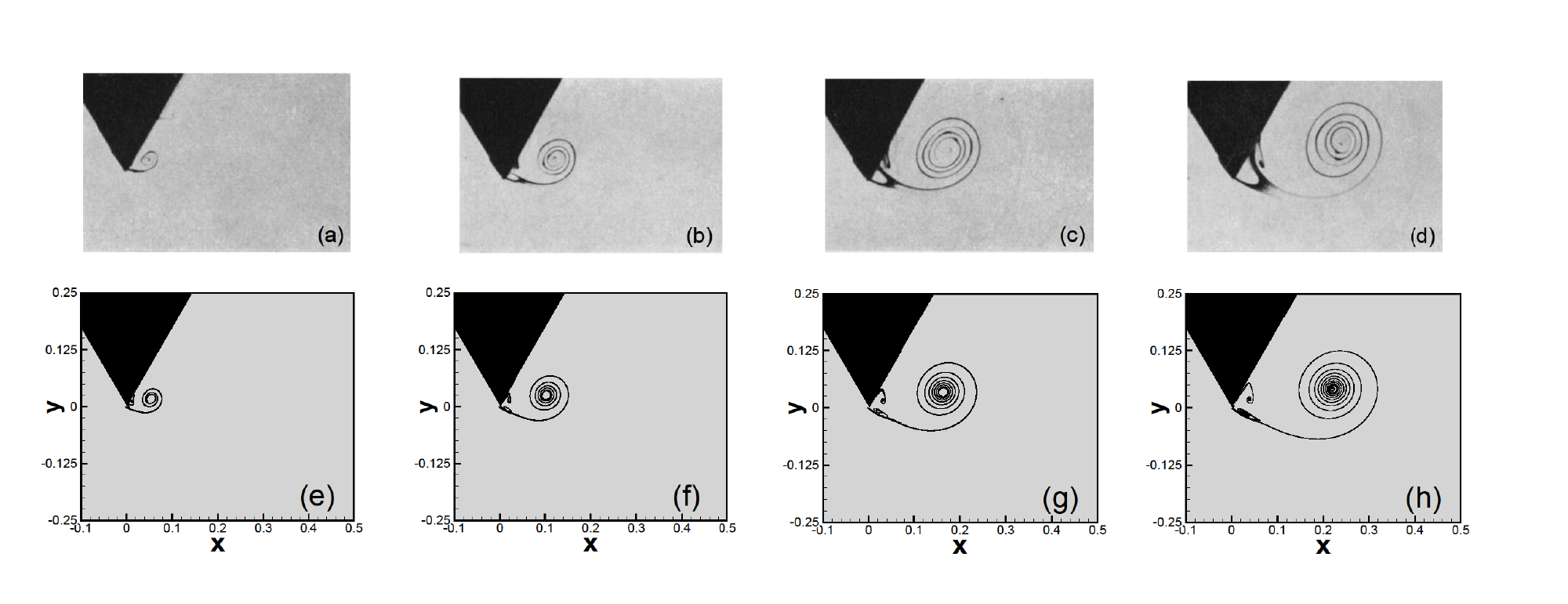}
\vspace{-0.75cm}
\caption{\sl { Streaklines for $Re_c=1560$ from  \cite{pullin1980} (a-d) and our computation (e-h) at instants $\tilde{t}=1s$ ($t=0.024803$),  $\tilde{t}=3s$ ($t=0.074409$), $\tilde{t}=5s$ ($t=0.124016$) and $\tilde{t}=7s$ ($t=0.173622$).}}
	\label{sk_wed_1560}
\end{figure}

	Computations were performed on grids of sizes $641 \times 161 $ $961 \times 241$, $1441 \times 361 $. For the first two grids, a uniform time step $\Delta t=10^{-4}$ was employed, while for the third, we used $\Delta t=5 \times 10^{-5}$. We carry out a grid-independence study by presenting the vorticity distribution along the wedge-wall at three time stations $t=0.2$, $0.3$ and $0.4$ on these three grids in figure \ref{grid_ind}(a)-(c). The overlapping of these graphs clearly indicates the grid-independence of the numerical computation.  Likewise, the vorticity contours plotted in figure \ref{grid_ind}(d) for the same grid sizes at tile $t=0.52$ after the stopping vortex is formed, leans towards the same conclusion.  Further, in order to validate our code, we present the streaklines of the experimental visualizations of \cite{pullin1980} for the earliest stage of the flow evolution side by side with the ones obtained from our simulations in figure \ref{sk_wed_1560}. The close proximity of the experimental and numerical results in capturing the flow characteristics such as the location of the primary and secondary vortex centers, their size and shape, instants of their occurrence clearly demonstrates the accuracy of the computational data.

	\section{Results and Discussion}\label{result}
     As mentioned in section \ref{intro}, the motivation of the current work comes from the experimental visualizations reported by \citet{pullin1980} in their start-stop lab experiments. Most of the previous numerical investigations (\cite{koumoutsakos1996,xu2006,xu2015,xu2016,kumar2020,kalita2023}) pertained to the starting vortex associated with the flow. Therefore, we embark on investigating the flow behind and rear side of the wedge when the incoming flow is stopped and how the deceleration applied to bring the incoming flow to a complete halt affects the flow pattern.

	\subsection{Flow development at the earliest stage}\label{fl_early}
 \begin{figure}[!H]
\begin{tabular}{cccc}
\hspace{-1.1cm}\epsfig{file=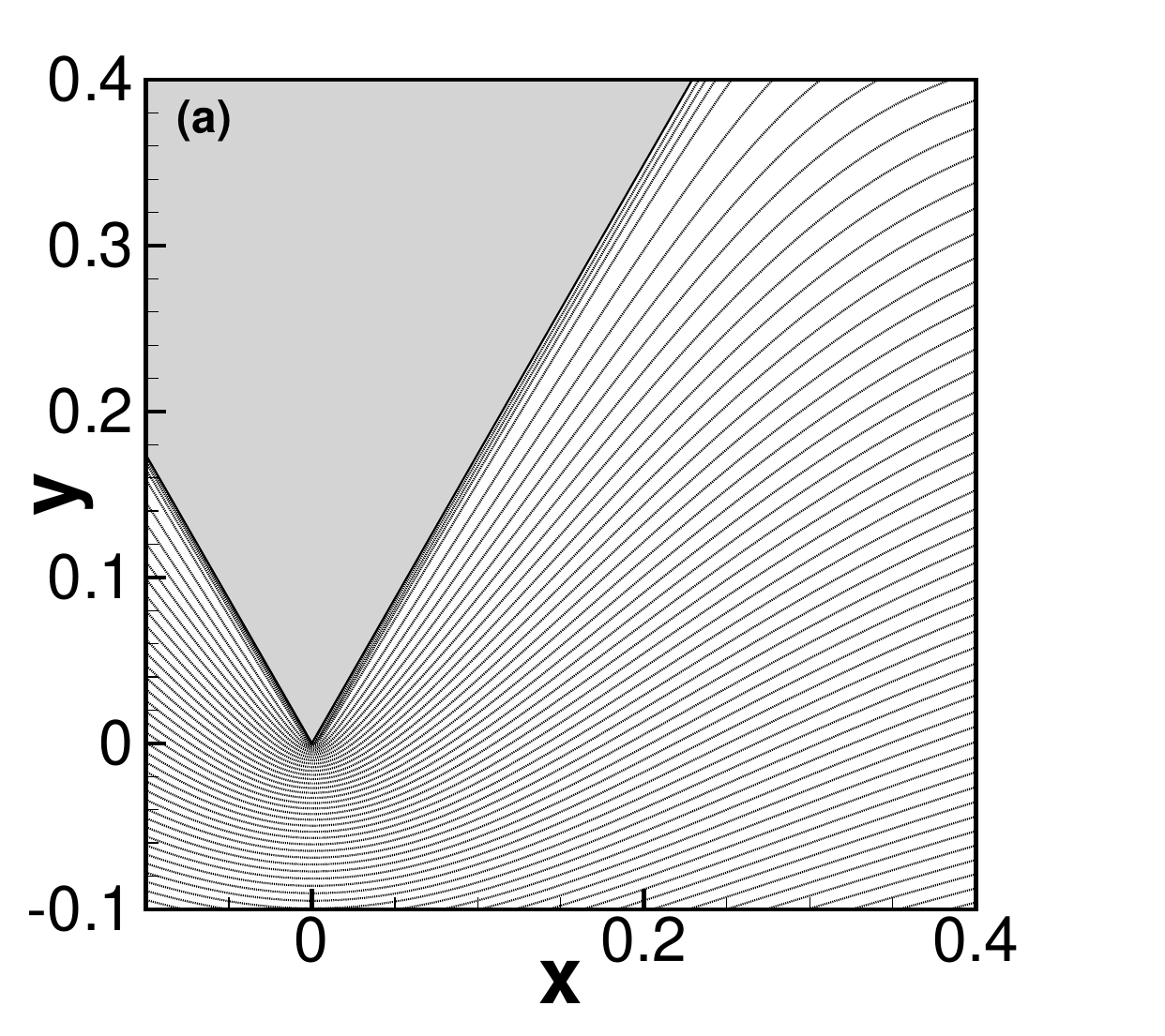,width=0.3\linewidth,clip=}
&
\hspace{-0.4cm}\epsfig{file=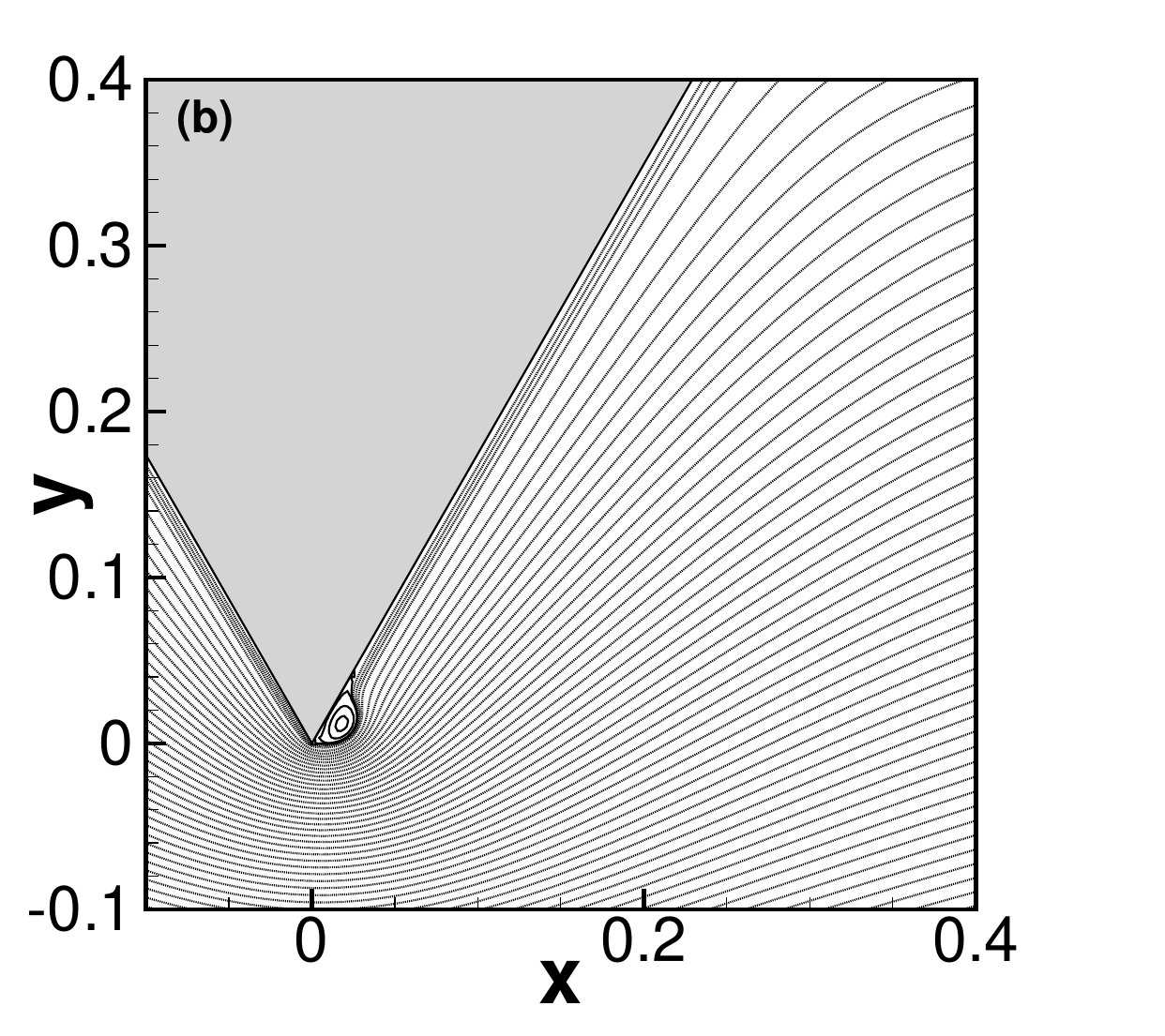,width=0.3\linewidth,clip=}
\&
\hspace{-0.4cm}\epsfig{file=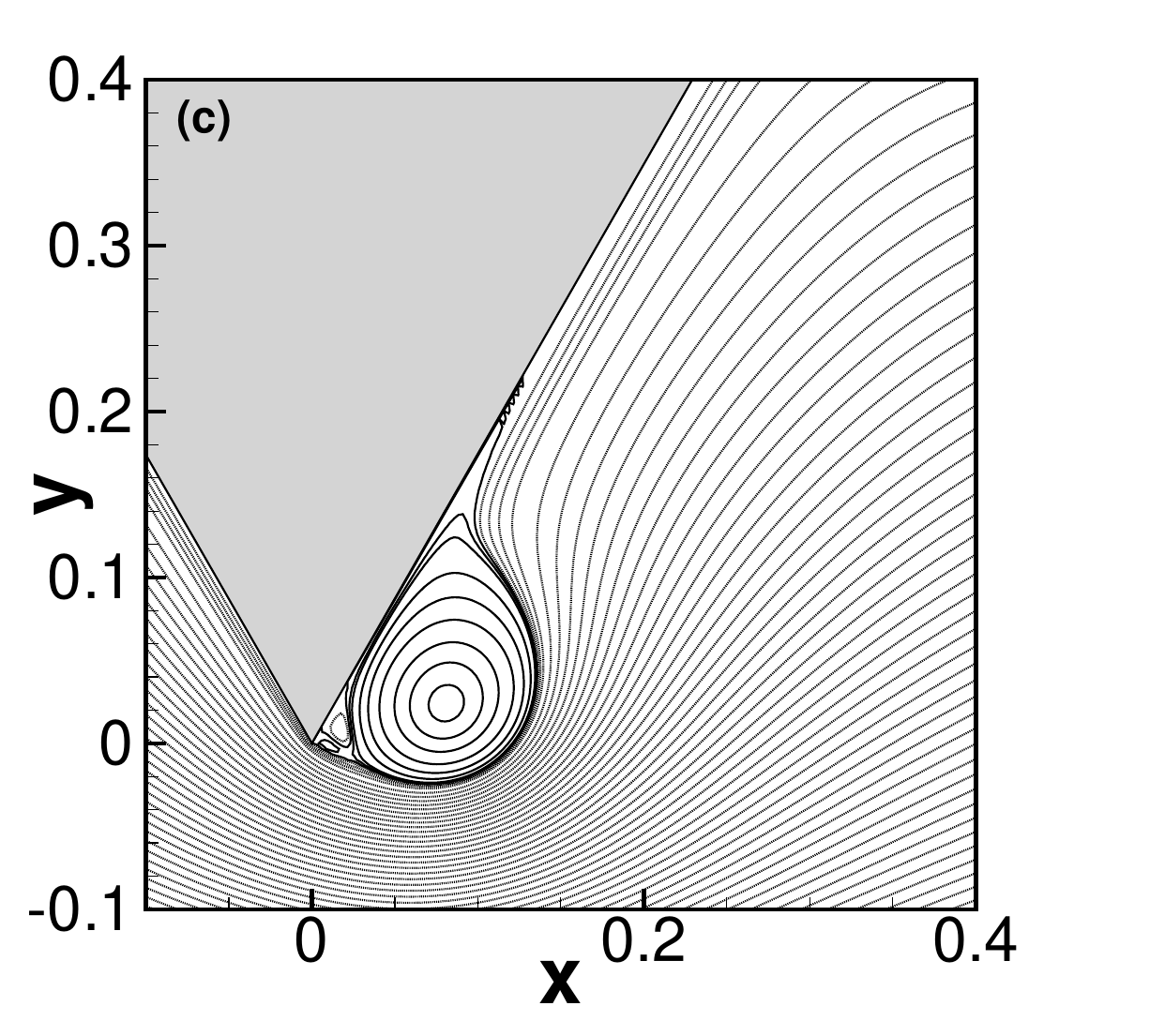,width=0.3\linewidth,clip=}
&
\hspace{-0.4cm}\epsfig{file=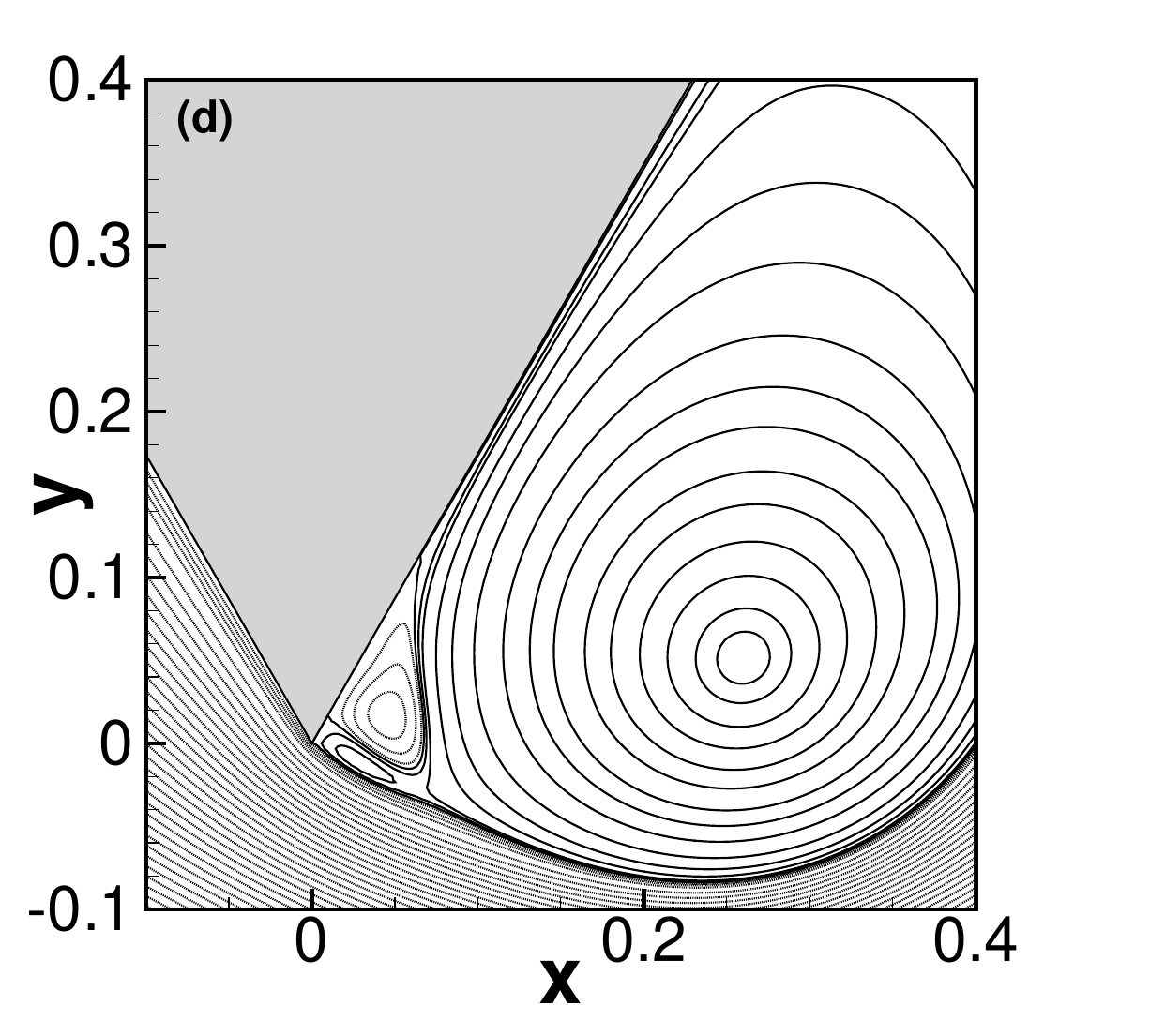,width=0.3\linewidth,clip=}
\end{tabular}
\begin{tabular}{cccc}
\hspace{-0.6cm}\epsfig{file=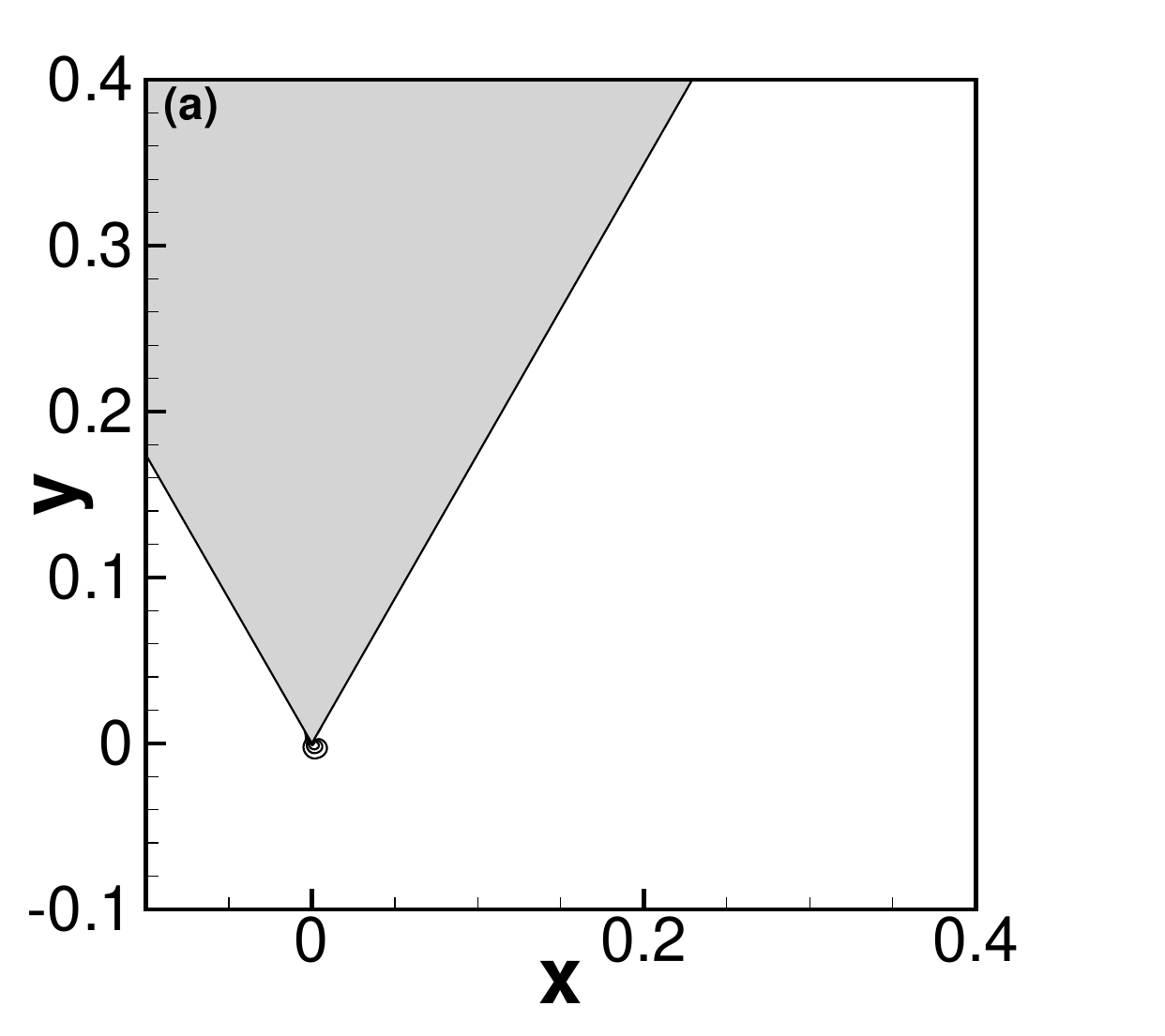,width=0.3\linewidth,clip=}
&
\hspace{-0.4cm}\epsfig{file=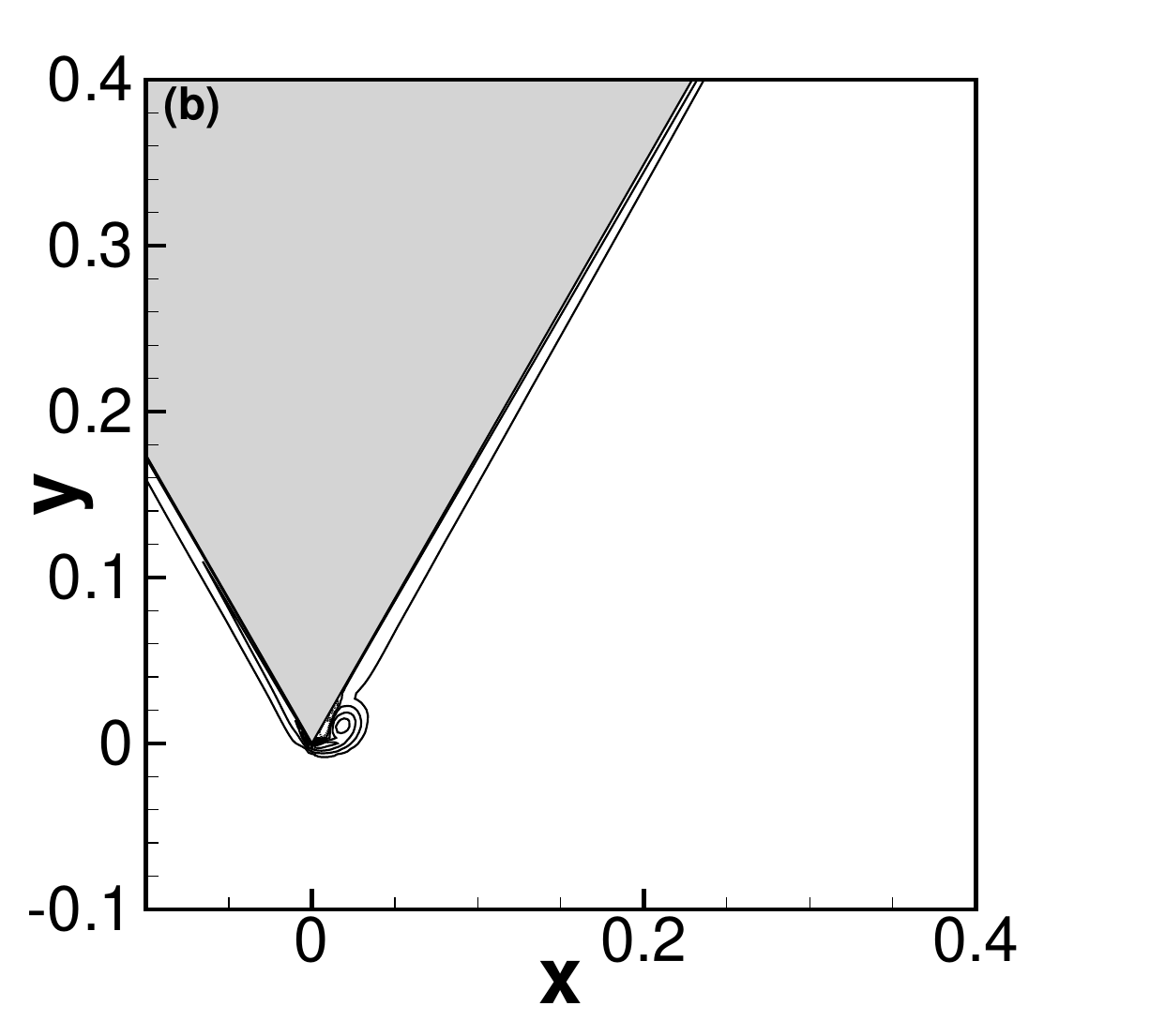,width=0.3\linewidth,clip=}
\&
\hspace{-0.4cm}\epsfig{file=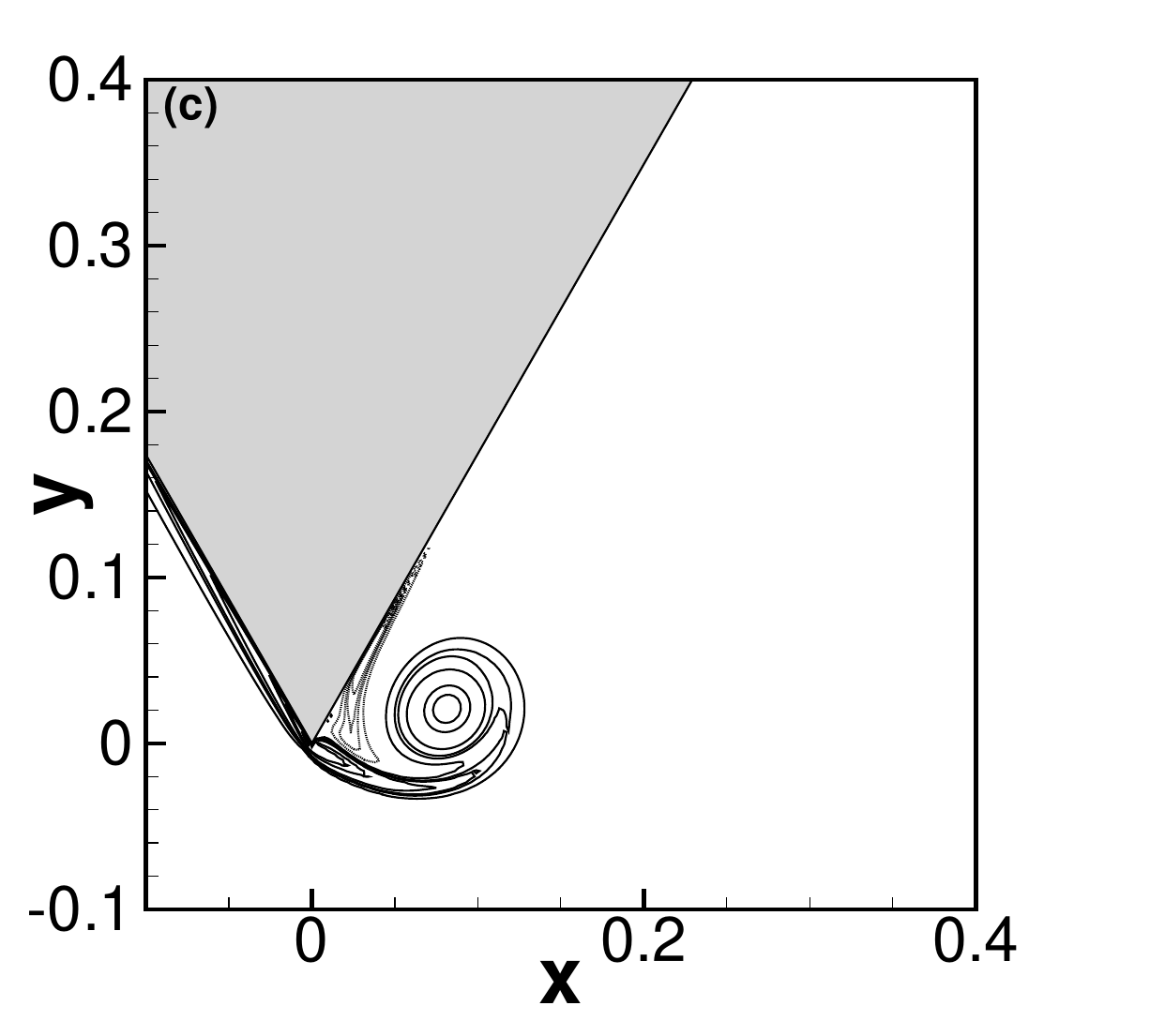,width=0.3\linewidth,clip=}
&
\hspace{-0.4cm}\epsfig{file=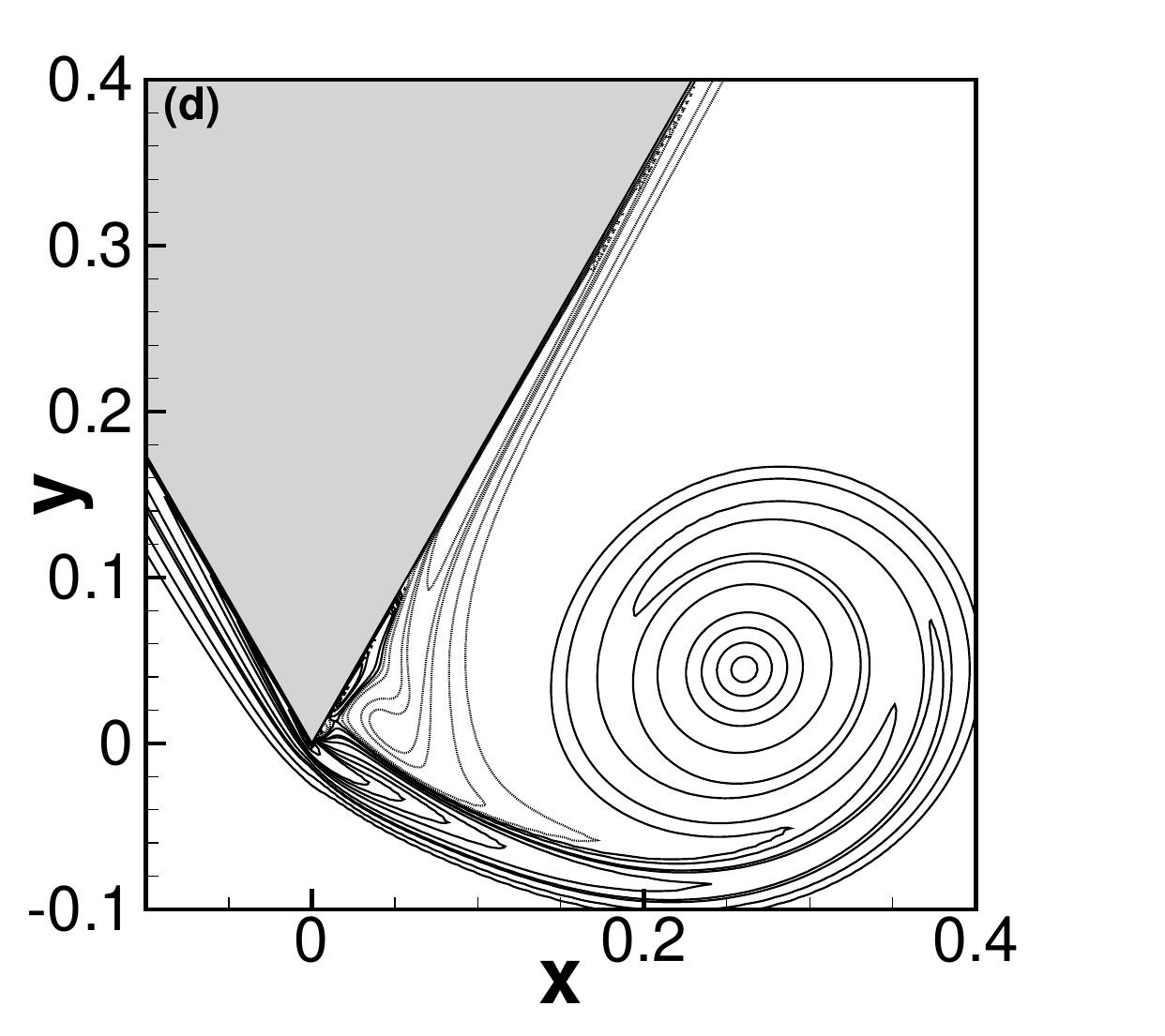,width=0.3\linewidth,clip=}
\end{tabular}
\caption{{Four stages of flow evolution for $Re_c=1560$ at non-dimensional time (a) $t=0.001$, (b) $t=0.01$, (c) $t=0.10$ and  (d) $t=0.40$: Top row, the streamlines and Bottom row, the vorticity contours. The positive contours are in solid lines and negative contours are in dotted lines.}}
\label{rayleigh}
\end{figure}
    The initial development of the flow is characterised by the presence of four stages of flow evolution documented by Luchini and Tognaccini \cite{luchini2002}, viz., the Rayleigh stage, Viscous stage, Self-similar inviscid stage and Vortex expulsion. These four stages are depicted in figures \ref{rayleigh}(a)-(d), where the top row represents the streamlines and the bottom one, the vorticity contours. At $t=0.001$,  vorticity and streamfunction contours of almost symmetrical shape around the tip is observed (figures \ref{rayleigh}(a)). all the vorticity values are positive at this stage, known as the Rayleigh stage duration of which culminates with the advent of the appearance of negative vorticity contours. They form within the starting vortex near the wall with a well-defined rotation center and bounding streamline as can be seen in figures figures \ref{rayleigh}(b) at time $t=0.01$. This is the viscous stage where the convective and diffusive terms contributes equally to the flow development, announcing the appearance of a well-defined vortex structure. After a while, the initial vortex separates from the wedge and the vorticity attains a local maximum at its core, which was absent earlier till that stage. This stage is termed as the self-similar stage (time $t=0.1$ figures \ref{rayleigh}(c)) where convection dominates over viscous diffusion. Then the starting vortex moves downstream, freeing itself from the self-similar growth and a second vortex is formed in the vicinity of the wedge-tip as in figure \ref{rayleigh}(d) at time $t=0.4$. This is vortex expulsion.

	\subsection{The stopping vortex}\label{stop_effect}
     Just while in the process of vortex expulsion, at the instant $t=0.435$, the incoming flow is decelerated (equivalent to reducing the piston velocity $v_p$) so as to reach a zero velocity value from unity through an interval of length $t_{dec}$. The variation of the incoming velocity over the time interval $[0,0.7]$ against several values of  $t_{dec}$ is plotted in figure \ref{velocity}. It would be revealed later on that the interval length $t_{dec}$ plays a significant role in the emergence, formation and the flow characteristics of the stopping vortex, and its subsequent effect on the evolution of the starting vortex.
	\begin{figure}
	\hspace{1.5cm}\centering\psfig{file=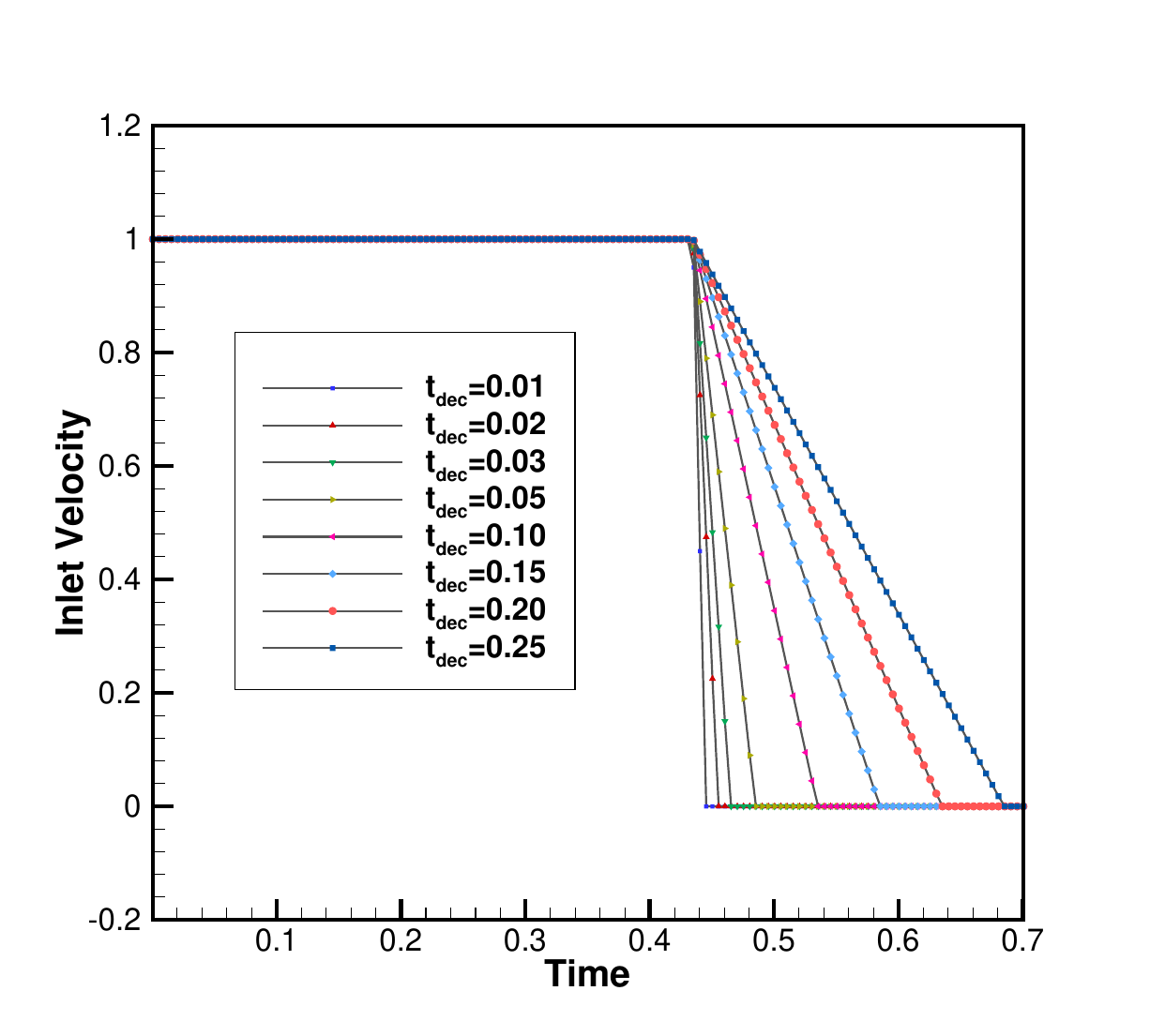,width=0.8\textwidth} 
	\caption{Velocity variation at the inlet for several deceleration time intervals of length $t_{dec}$ depicting the cases with and without stopping vortex  for $Re=1560$ after the impulsive stop.}
	\label{velocity}
    \end{figure}

\begin{figure}
 \includegraphics[width=5.5in]{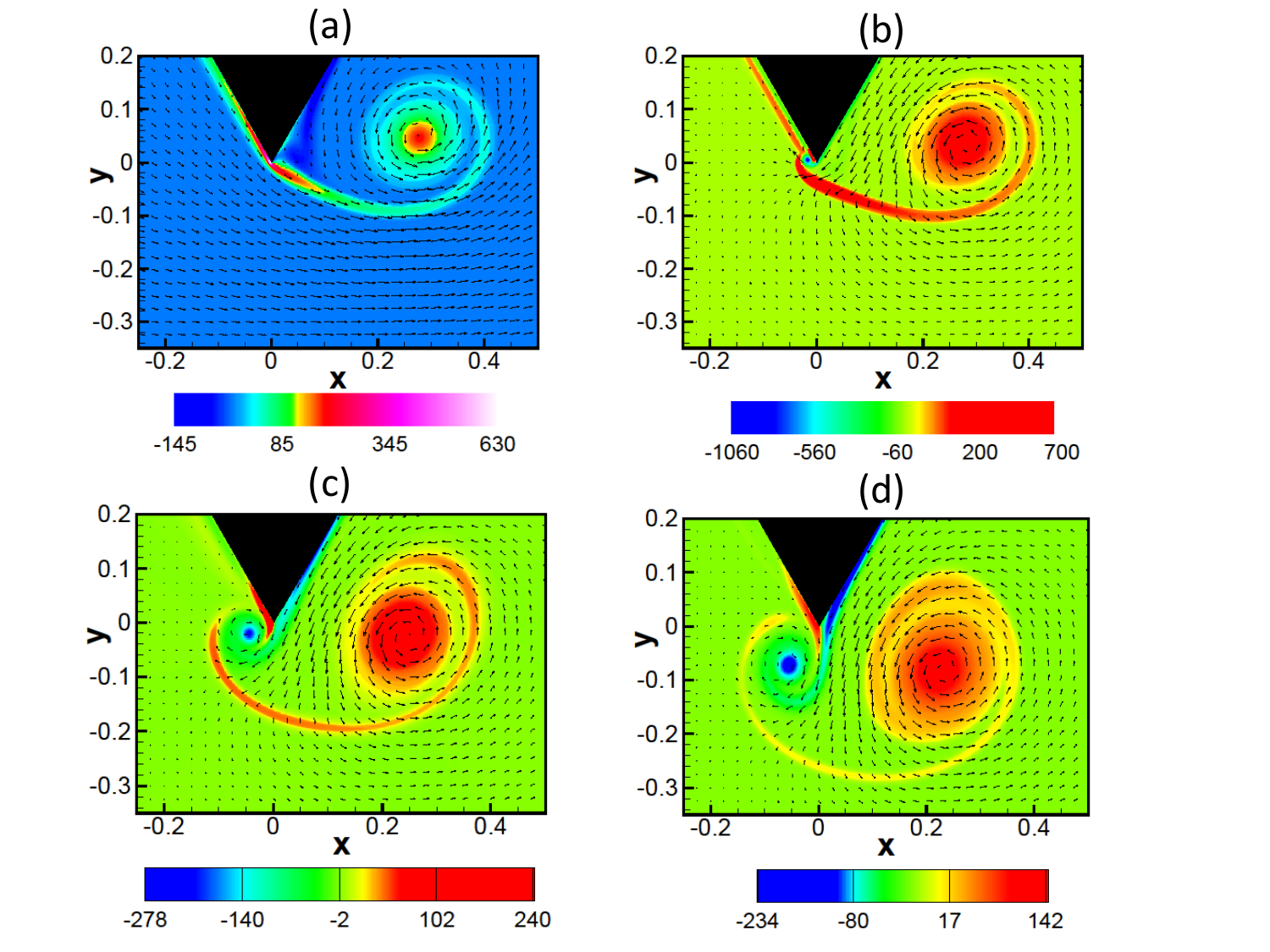}
\vspace{-.55cm}
	\caption{Evolution of vorticity contours for the starting and stopping flow before and after the impulsive stop: (a) $t=0.43$, (b) $t=0.45$,  (c) $t=0.55$ and (d) $t=0.65$. The deceleration of the flow starts at $t=0.435$ and the flow stops at $t=0.437$.}
	\label{ev_vort}
	
\end{figure}

    Once the flow stops, another vortex emanates from the tip of the wedge moving in the direction opposite  to the starting vortex. The roll-up velocity of the starting vortex induces the generation of this secondary vortex of opposed circulation at the wedge-tip. While the vorticity inside the starting vortex is positive having an anticlockwise rotation, the orientation of rotation is clockwise for the stopping vortex endowed with negative vorticity. This can be seen from the evolution of the vortices shown from just on the edge of process of deceleration in figures \ref{ev_vort}(a)-(d). Here the evolution of the vorticity contours along with the velocity vectors can be seen at instants $t=0.43$, $0.45$, $0.55$ and $0.65$ respectively.  One can clearly see that both the stopping and starting vortices follow a downward movement after the incoming flow is stopped (see figures \ref{velocity} and \ref{core_omega}(b) also). 
	
	\subsubsection{Schematics of the vortex center, radius and circulation}
	\begin{figure}
	\centering\psfig{file=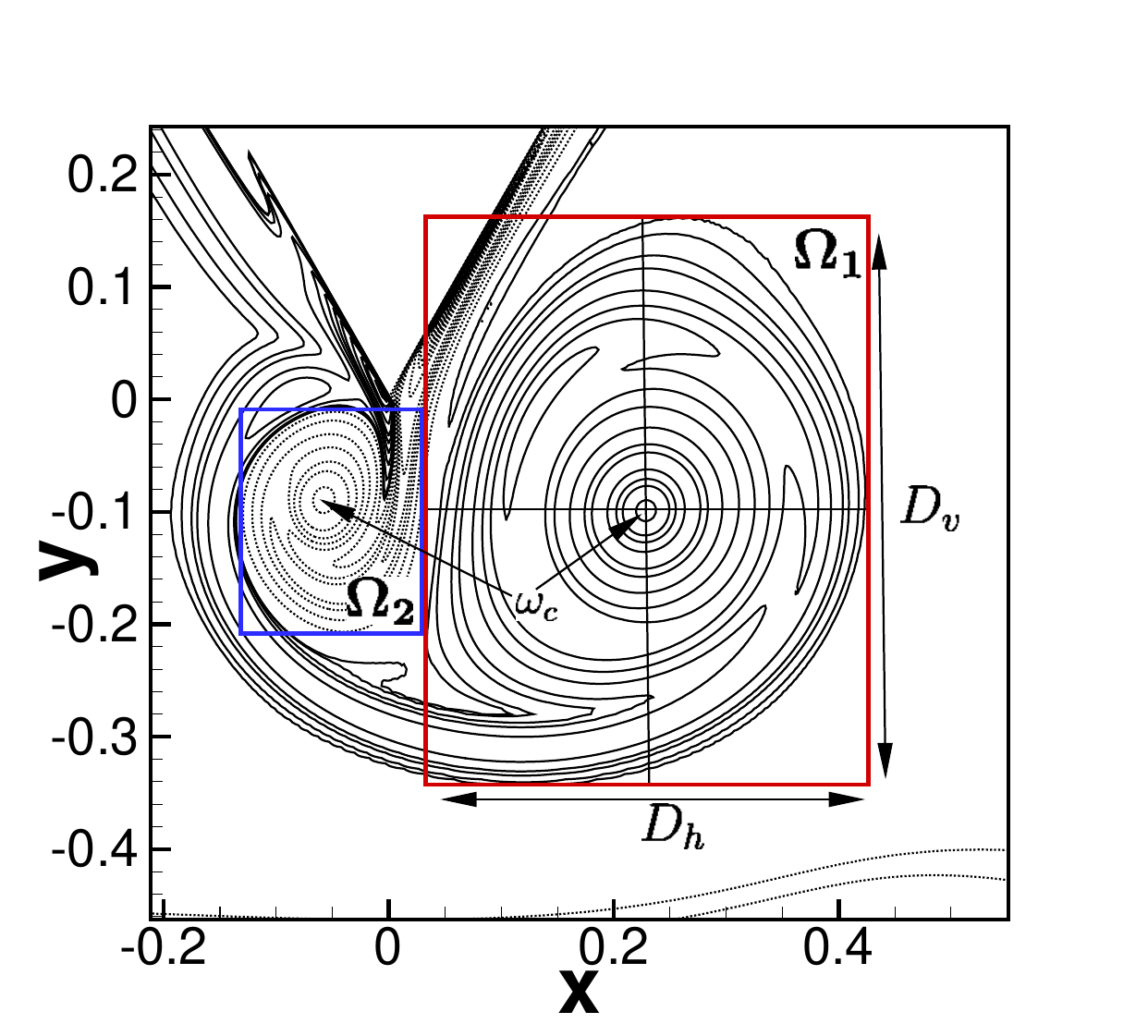,width=0.8\textwidth}
	\caption{{Schematic of vortex center, radius and circulation.} }
	\label{sc_center}
\end{figure}
	We present the schematics of the vortex centre rand its radius in figure \ref{sc_center}. For the starting vortex, the vortex center is defined as the point in the computational domain having local vorticity maximum $\omega_M$ inside the primary vortex in the leeward side of the wedge, where the vortcity contours are depicted as solid curves.  The radius $R_\omega$ is defined as $\displaystyle \frac{D_h+D_v}{4}$, where $D_h$ and $D_v$ are the lengths of the horizontal and the vertical sides of the rectangular region $\Omega_1$ (shown in red) inscribing the starting vortex completely as shown in figure \ref{sc_center}. The extent of $\Omega_1$ is determined by an algorithm which searches for positive vorticities in the horizontal and vertical directions about $\omega_M$. Once it encounters a negative vorticity, the search is stopped. Likewise, for the stopping vortex, the vortex center is the point of local minimum in the windward side of the wedge, where the vortcity are depicted by dotted curves in the figure. The radius is computed in a similar manner from the rectangular region $\Omega_2$ which is determined by an analogous algorithm as the one mentioned earlier, except that it now searches for negative vorticities about the vortex center. We define the circulation $\Gamma$, which is a measure of the strength of the vortex at any instant $t$ as $\displaystyle \Gamma (t)=\int_{\Omega_1}\omega(x,y,t)dA,\; \omega>0$ for the starting vortex and  $\displaystyle \Gamma (t)=\int_{\Omega_2}\omega(x,y,t)dA,\; \omega<0$ for the stopping vortex, where $\Omega_1$ and $\Omega_2$ are the rectangular domains as shown in figure \ref{sc_center}.
	
	\subsubsection{Effect of the length $t_{dec}$ of deceleration time interval on flow}
     The length $t_{dec}$ of deceleration time interval plays a crucial role in the development of the overall flow after the incoming flow comes to a complete halt. In figure \ref{velocity}, we show the trajectories of the vortex centers for different values of $t_{dec}$. The trajectories of the starting vortex corresponds to the time interval $[0,0.7]$. Till deceleration is applied, the centers of the starting vortex move in the downstream direction almost along a slightly elevated straight line. Once the flow is stopped, the forward movement slows down and the vortex centers begin to move downwards. Smaller is the length $t_{dec}$, faster is the downward movement and longer is the distance traversed. On the other hand, a larger value of $t_{dec}$ aids the forward movement of rolled-up starting vortex as can be seen from the extreme cases of $t_{dec}=0.01$ and $t_{dec}=0.20$ shown in the figure \ref{velocity}.  
	\begin{figure}
	\centering\psfig{file=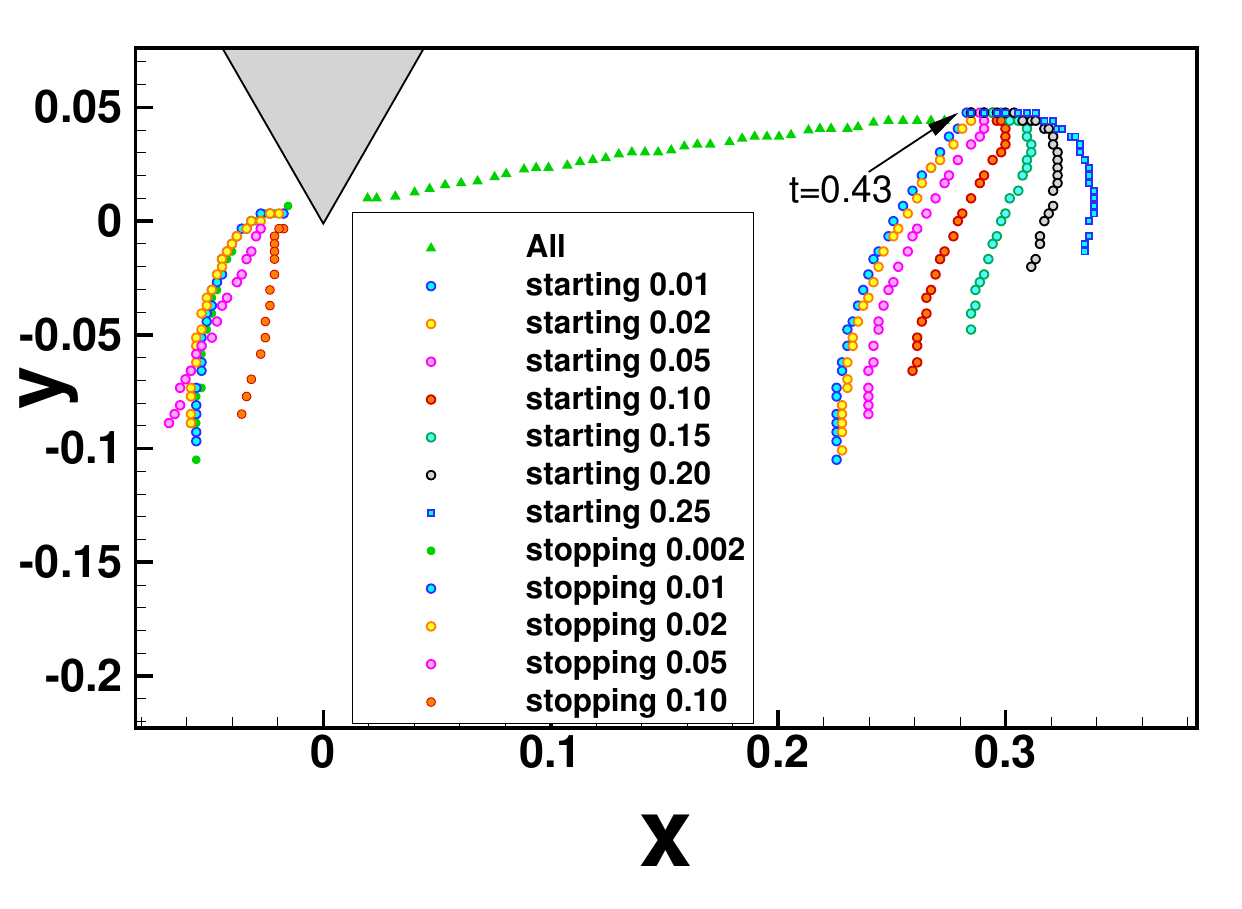,width=\textwidth} 
	\caption{ Trajectories of the starting and stopping vortex centers for different time intervals of deceleration till $t=0.70$. On the left of the wedge tip are the trajectories for the stopping vortex for $t>0.435$ and on right, the starting vortex center trajectories that change track after $t=0.435$.}
	\label{velocity}
\end{figure}

\begin{figure}
	\begin{center}
	\includegraphics[width=0.445\textwidth]{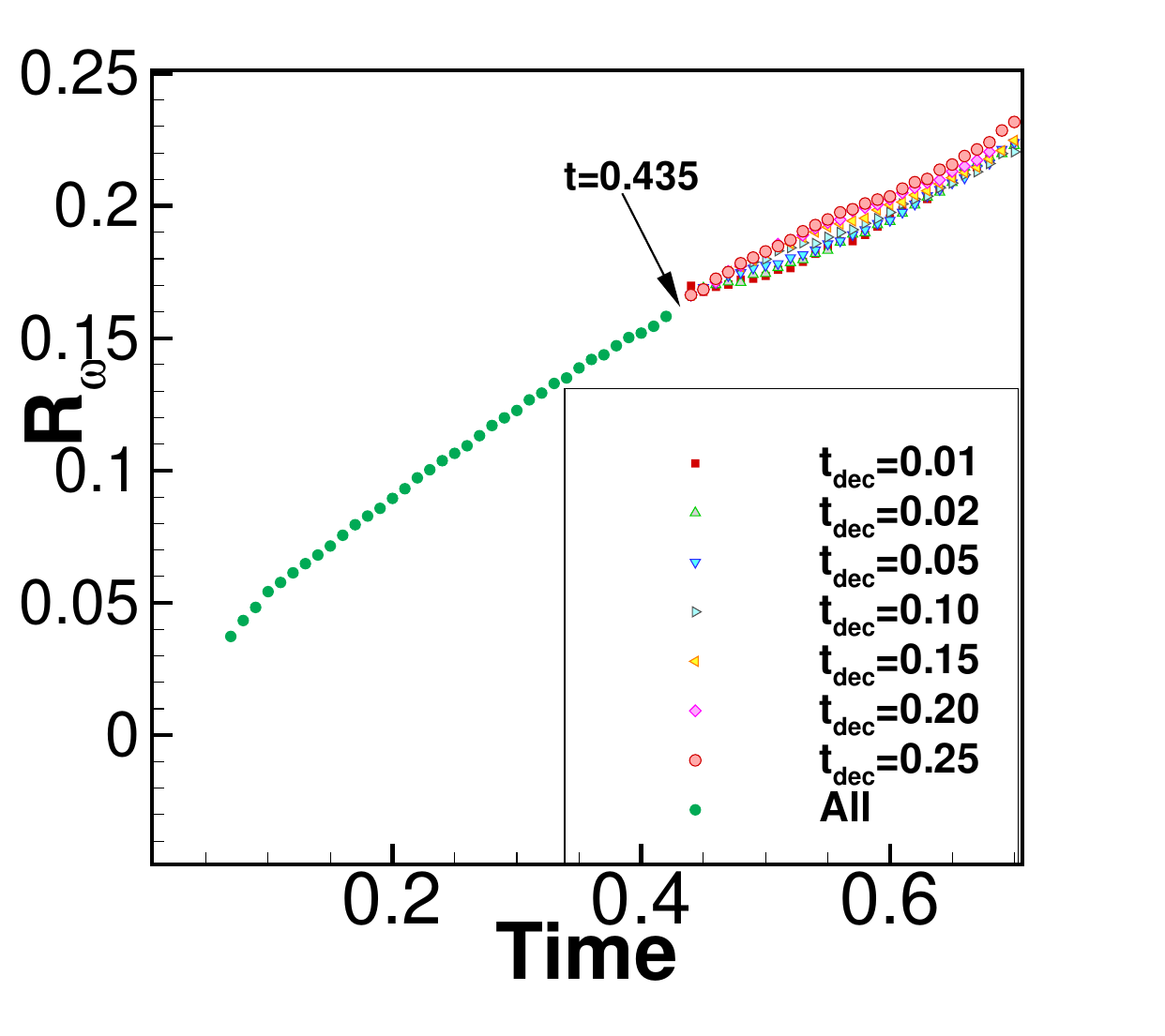}
	{(a)}
	\includegraphics[width=0.445\textwidth]{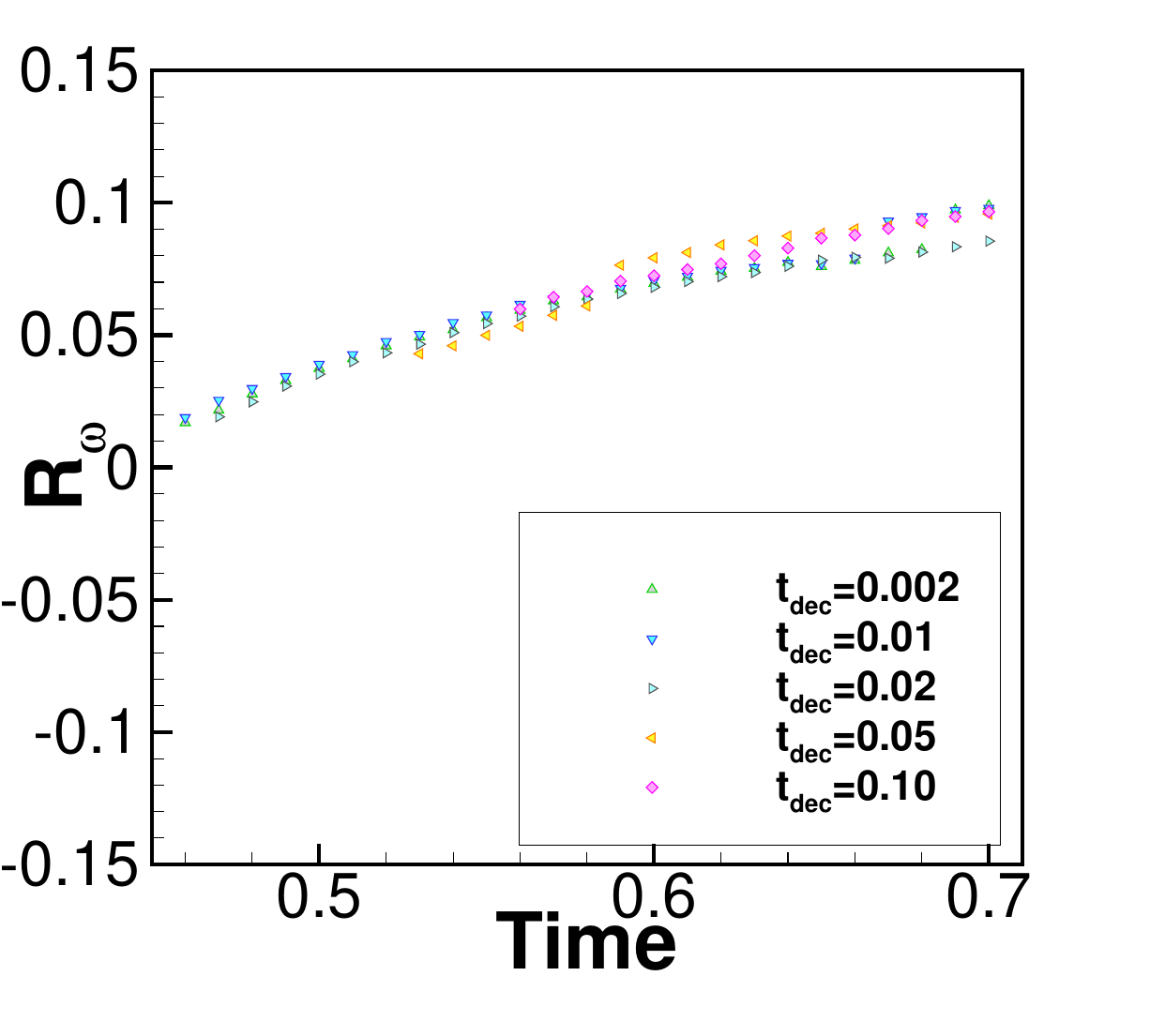}
	{(b)}
	\caption{ Evolution of radii $R_{\omega}$ of the (a) starting and (b) the stopping vortices for different time intervals of deceleration.}
	\label{rad}
	\end{center}
\end{figure}

\begin{figure}
	\begin{center}
	\includegraphics[width=0.445\textwidth]{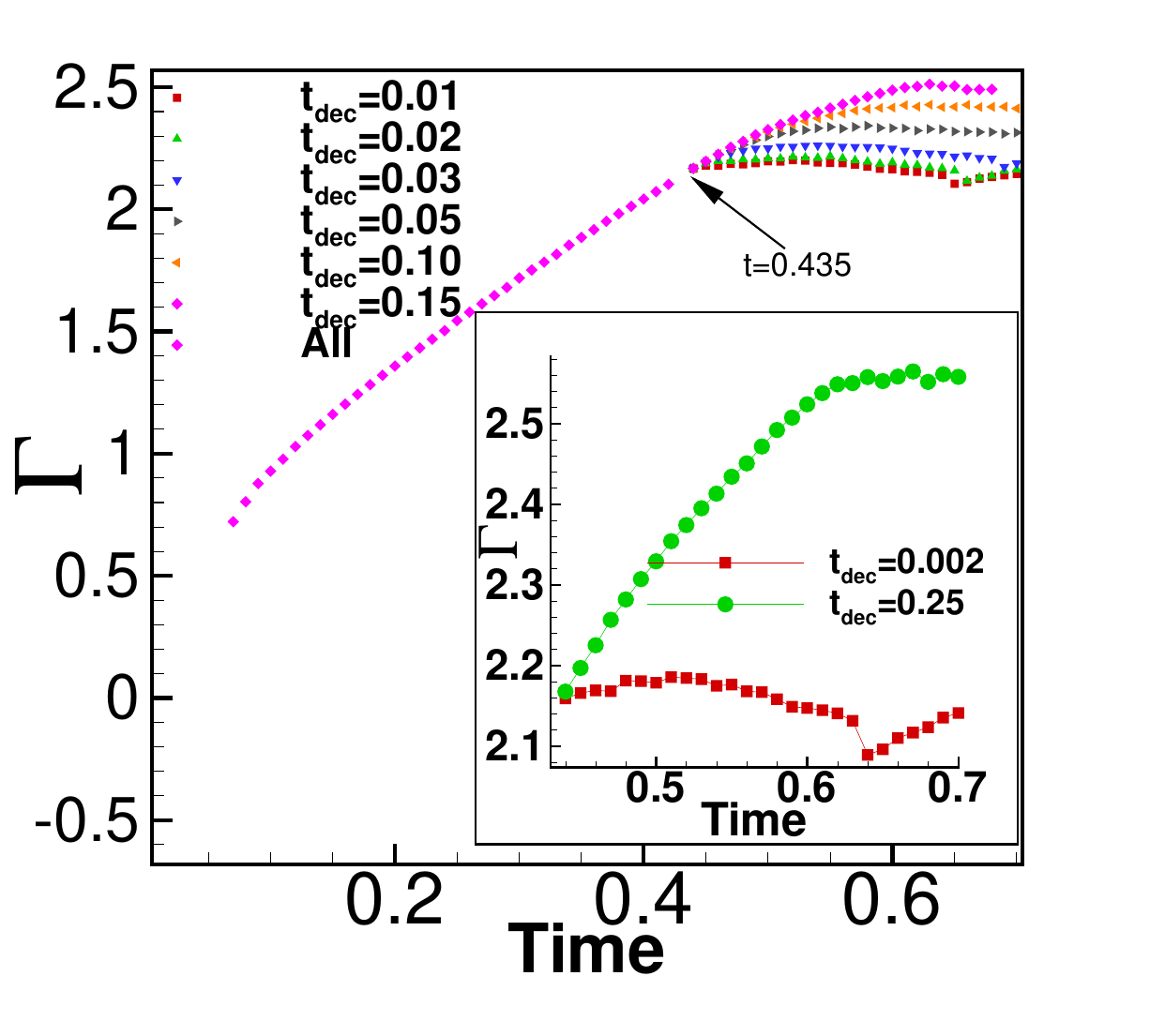}
	{(a)}
	\includegraphics[width=0.445\textwidth]{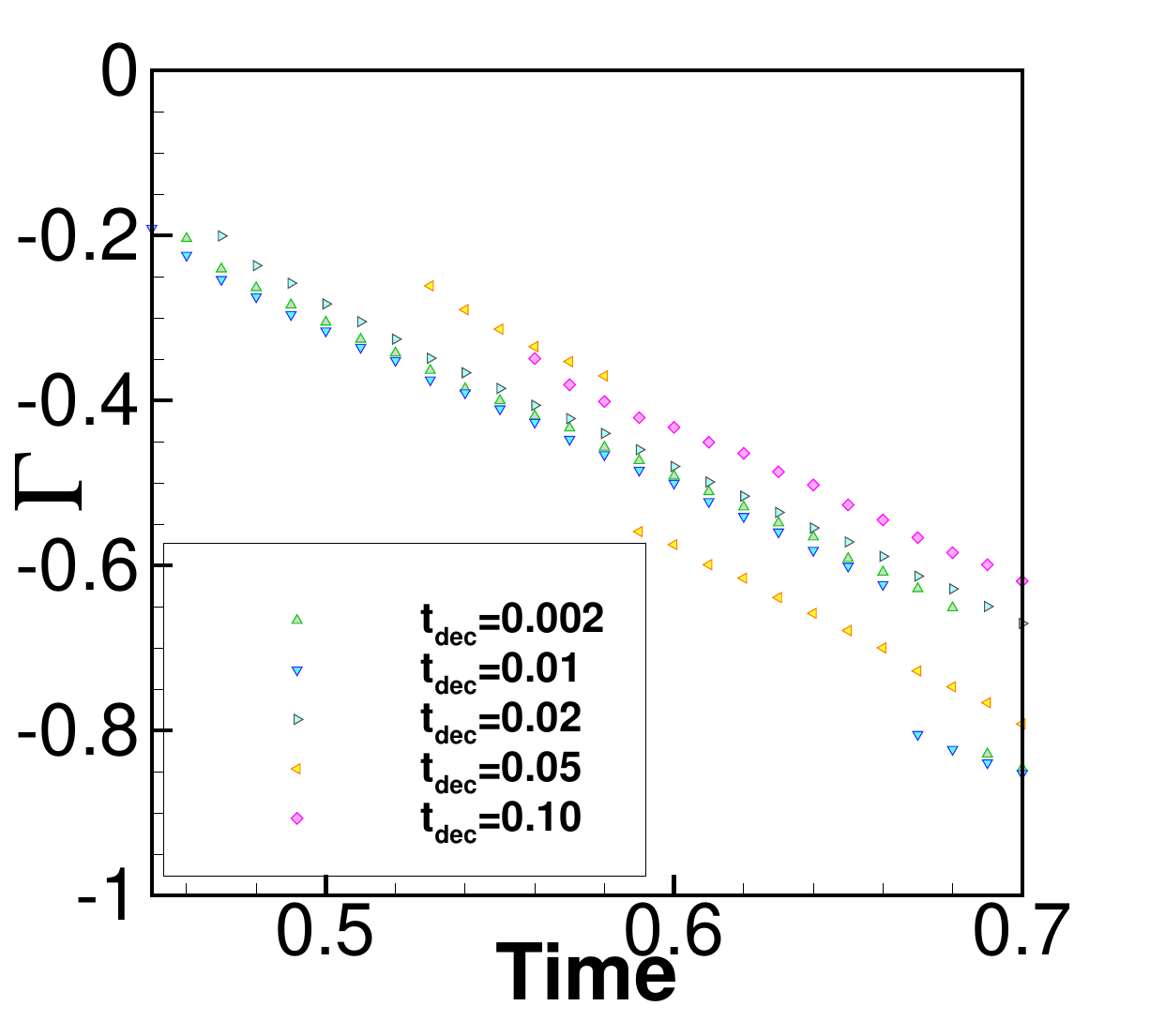}
	{(b)}
	\caption{ Evolution of circulation $\Gamma$ of the (a) starting and (b) the stopping vortices for different time intervals of deceleration.}
	\label{circulation}
	\end{center}
\end{figure}

     Once the stopping vortex starts appearing, its starts moving along a trajectory in the south western direction of the wedge-tip. Here the trajectories correspond to the interval $[0.435,0.7]$. Note that,  the local minimum at the core of the vortex is attained only after the incoming flow comes to a complete halt.  Once again, as in the case of the starting vortex, smaller the value of $t_{dec}$, larger is the distance traversed by the stopping vortex. A higher value of $t_{dec}$ restricts the vortex centers in the vicinity of the wedge-tip as can be seen from figure \ref{velocity}. The length of the time interval also affects the size of the stopping vortex as can be seen from the evolution of the radii of the stopping vortices in figure \ref{rad}(b). A shorter interval indicating a rapid deceleration facilitates a rapid growth, while a longer one drastically halts their growth. On the other hand, the formation of the stopping vortex slows down the growth rate of the starting vortex (figure \ref{rad}(a)). From the close-up view of the extreme cases considered here in the inbox of figure \ref{rad}(a), viz., $t_{dec}=0.25$ and $t_{dec}=0.002$, one can see that while a larger interval is seen to least affect the growth rate, for a shorter one, its effect is much more prominent as can be seen from the two extreme cases considered in this study. The observation for the circulation (in magnitude) is analogous to that of the radius for both the starting and stopping vortices as can be seen from figures  \ref{circulation}(a)-(b). For the stopping vortex, the shed circulation increases almost linearly during the formation period (\cite{wakelin1997}). However, when the incoming flow is stopped at $t=0.435$, the fall in its value is not as sharp as observed in \cite{wakelin1997} in the case of accelerated flow past a circular orifice. A possible cause for this could be the uniform nature of the incoming flow.


\subsubsection{Criterion for a clean stopping vortex}
\begin{figure}
	\begin{center}
	\includegraphics[width=0.445\textwidth]{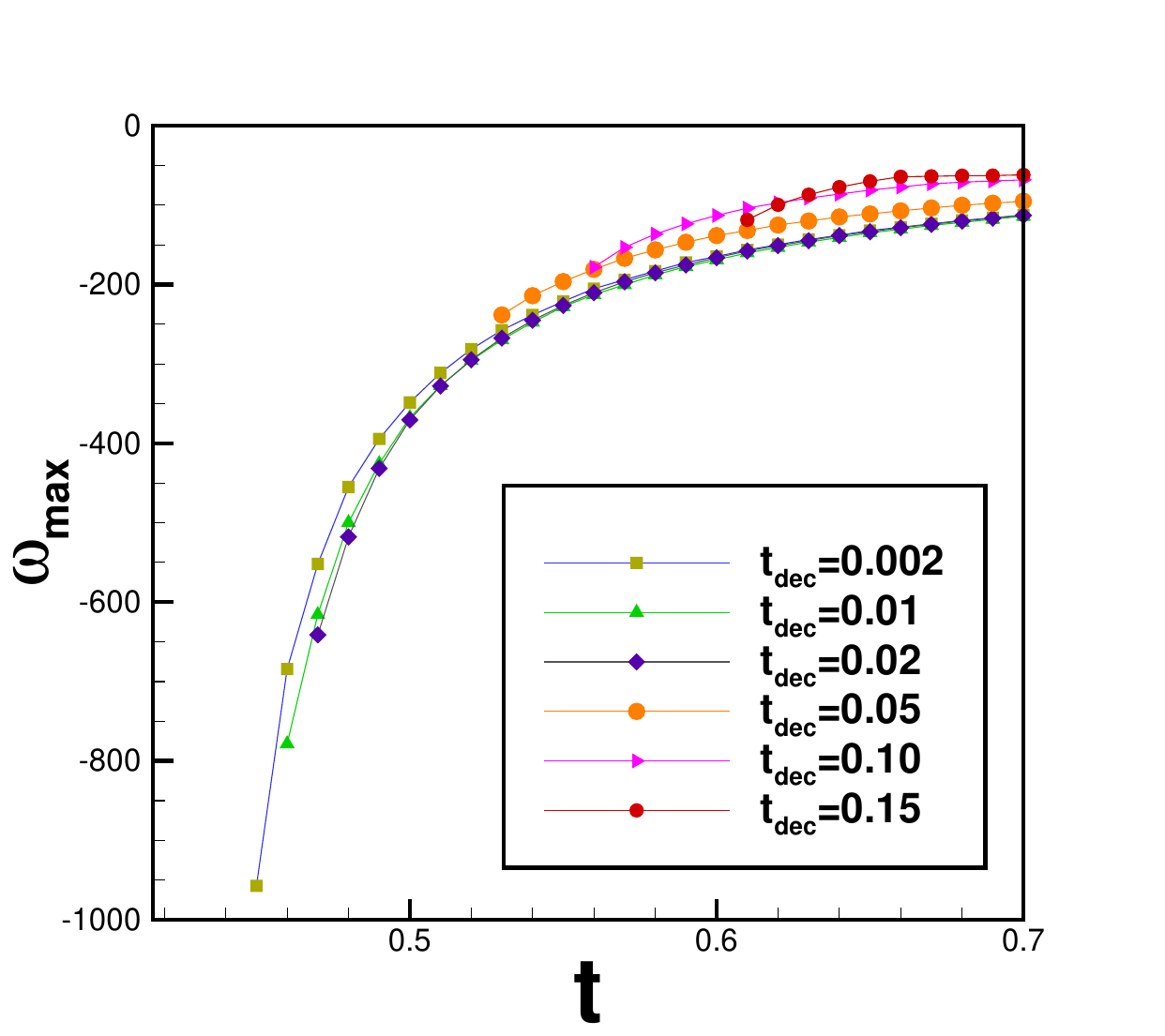}
	{(a)}
	\includegraphics[width=0.445\textwidth]{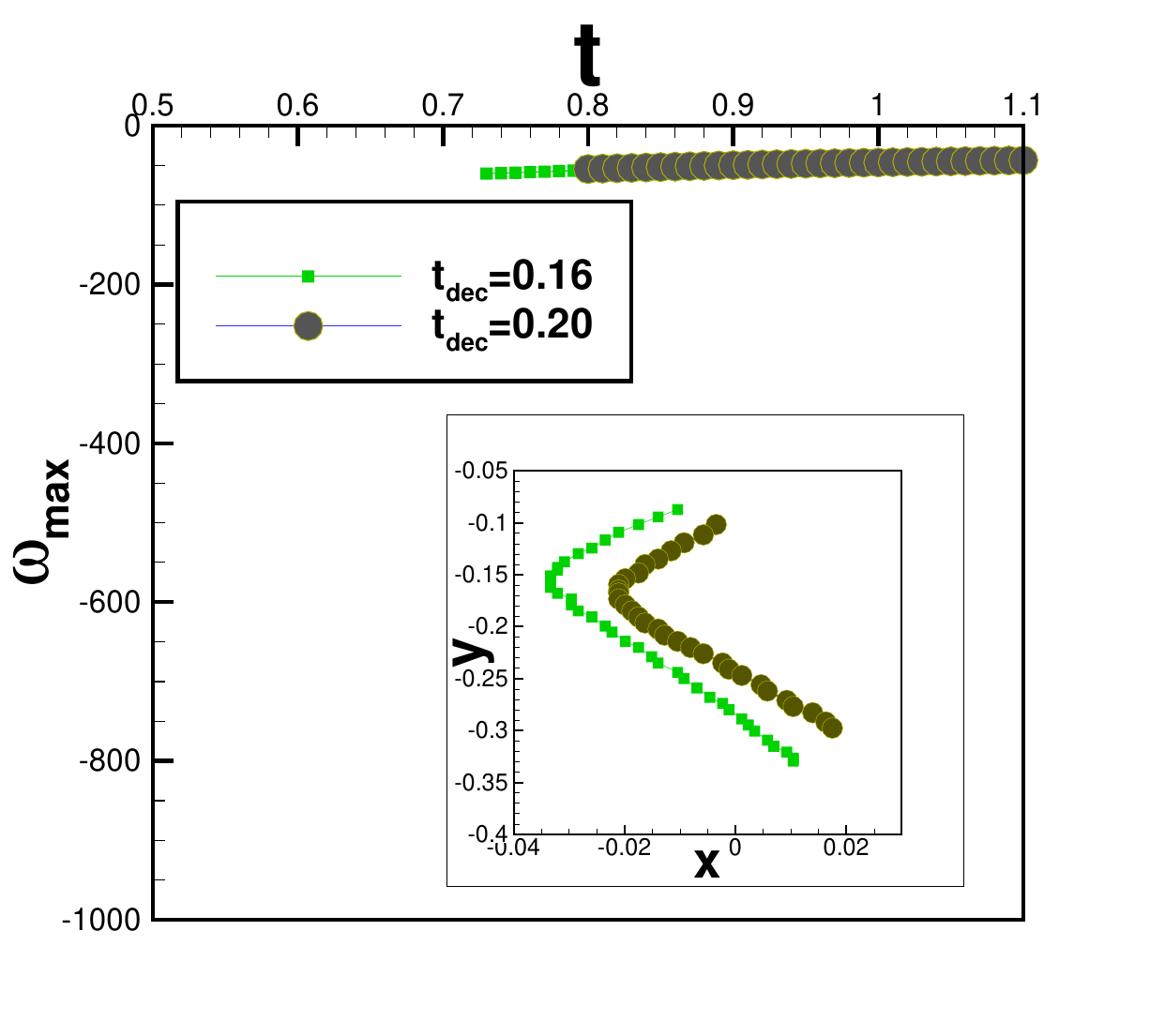}
	{(b)}
	\caption{ Evolution of (a) the vorticity minimum at the vortex center for the strong stopping vortices and (b) the vorticity minimum at the vortex center along with the trajectories of the vortex centers for the weak stopping vortex for different time intervals of deceleration.}
	\label{core_omega}
	\end{center}
\end{figure}

As depicted in section \ref{fl_early}, the early part of the starting vortex vortex is characterised by the viscous stage followed by the self-similar stage. Likewise, for the stopping vortex, the formation of a clean vortex must be  synonymous with the attainment of a local minimum by the vorticity at the core of the vortex after the separation from the wedge. However, a clear vortex core is not observed for the stopping vortex immediately after the streamlines separating from the wedge for all the cases. For extremely small $t_{dec}$, it is observed almost instantly after the incoming flow coming to a complete halt. The magnitude of vorticity at the vortex center is very high, indicating the strength of the vortex as can be seen from figure \ref{core_omega}(a), where the time history of the vorticity values at the vortex center is shown till $t=0.70$ for gradually increasing values of  $t_{dec}$ ranging from $0.002$ to $0.15$. On the other hand, as $t_{dec}$ increases, the appearance of the stopping vortex is delayed along with a gradual increase in its strength. In figure \ref{core_omega}(b), the time history of the vorticity values at the vortex center is shown till $t=1.10$ for $t_{dec}=0.16$ and $0.20$ along with the trajectories of the vortex centers at the inset. From figures \ref{core_omega}(a)-(b), one can clearly see that for $t_{dec}=0.15$, the local minimum of vorticity is visible from $t\approx 0.61$ onwards,  while for $t_{dec}=0.16$, it starts appearing from $t\approx 0.73$, indicating a distinct jump. Figure \ref{core_omega}(b) also indicates that there is a drastic reduction in the strength of the vortices. Next, we would define the criterion for the formation of a clean stopping vortex based on a model previously used by \cite{das2017}, which would reveal the consistency of our observation with theirs.

Contrary to the accelerated incoming flow in their laboratory experiment of \cite{das2017}, the current numerical experiment is concerned with a uniform incoming flow. Once the incoming flow decelerates, a reverse flow occurs in the left neighbourhood of the wedge-tip, which causes the entrained fluid motion from below to roll up and form the stopping vortex. Therefore, the vortex formation process must be related to the rapidity at which the incoming flow decelerates to zero velocity.  \cite{das2017} have quantified this rapidity as the fractional impulse imparted during the deceleration time.

\begin{figure}
\includegraphics[width=5.5in]{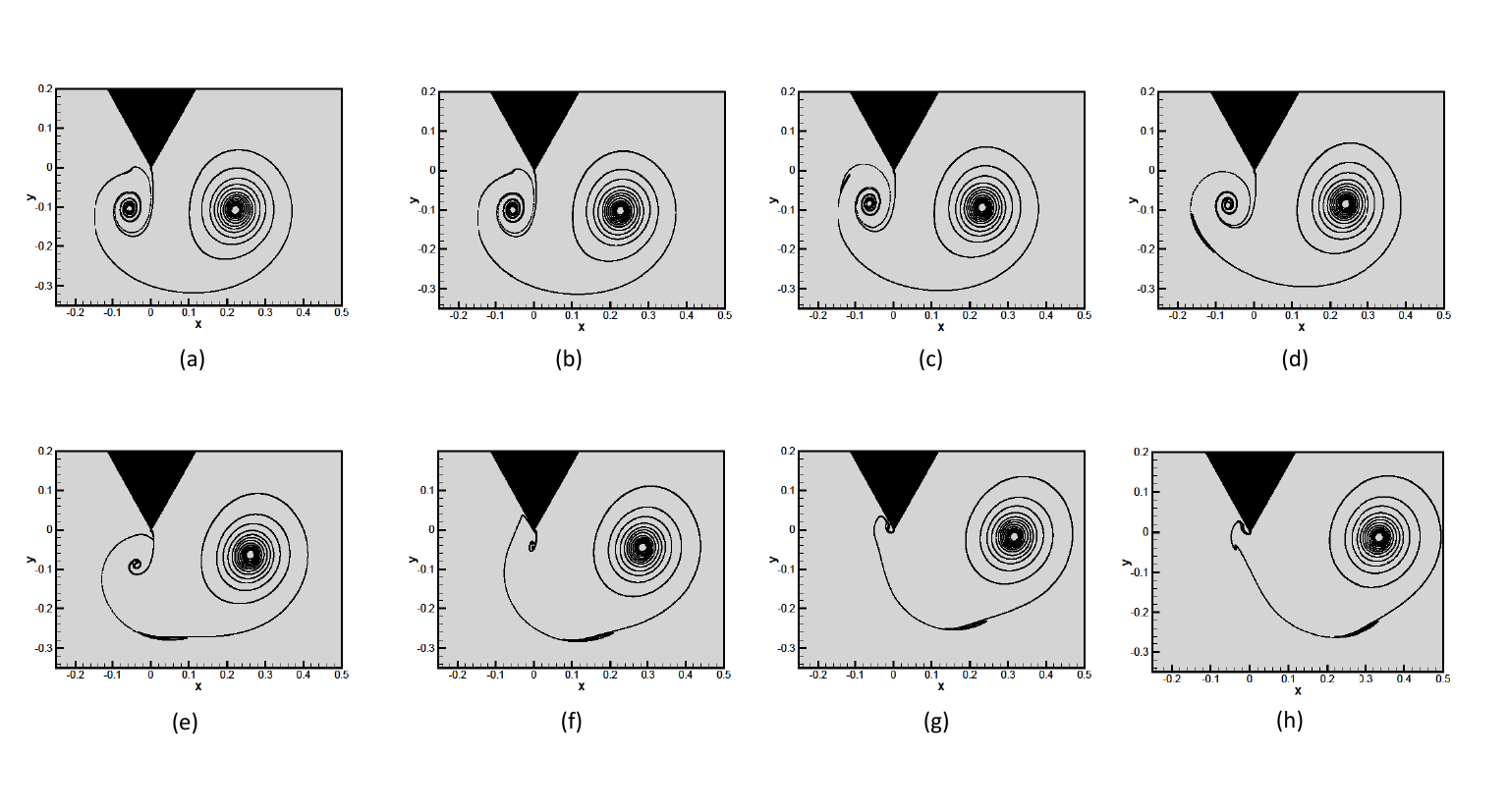}
\vspace{-.55cm}
\caption{\sl {Streaklines at time $t=0.70$ for $Re_c=1560$ after the impulsive stop at time $t=0.435+t_{dec}$ where $t_{dec}$ is the length of the time interval over which the inlet velocity reduces to zero from a unit velocity. The deceleration starts at time $t_f=0.435$. Top: $t_{dec}= {\rm (a)} 0.002,\; {\rm (b)}0.01,\;{\rm (c)}0.03\;{\rm and}\;{\rm (d)}0.05$ (from left to right) and Bottom: $t_{dec}={\rm (e)}0.10,\; {\rm (f)}0.15,\;{\rm (g)}0.20\;{\rm and}\;{\rm (h)}0.25$ (from left to right).}}
	\label{dec_gap}
\end{figure}

Let $I_U$ be the impulse imparted by the uniform incoming flow during the time interval $[0,t_f]$ and $I_T$ be the total impulse over the time interval $[0,t_f+t_{dec}]$, where the impulse $I$ over a certain interval is defined as $\displaystyle I=\int u^2 \d t $, $u$ being the inlet velocity. From our numerical experiments, we have seen that if the ratio $\displaystyle R_{imp}=\frac{I_U}{I_T}<0.90$ approximately, the stopping vortex either does not form or is extremely weak. 
Evidence of this can be clearly seen from figure \ref{dec_gap} where streaklines at time $t=0.7$ are shown for eight different values of  $t_{dec}$. While a clear vortex can be seen for $t_{dec}$ ranging from $0.002$ to $0.1$ (figures \ref{dec_gap}(a)-(e)), the stopping vortex is extremely weak or non-existent for $t_{dec} \geq 0.15$ (figures \ref{dec_gap}(f)-(h)). Note that $t_{dec} = 0.15$ corresponds to the value $R_{imp}=0.896907$. The strength of the stopping vortices could also be gauged from more rounds of rotations performed by the stopping vortices in the top row than the bottom one in figure \ref{dec_gap}, which is consistent with the vorticity values at the centers in figures \ref{core_omega}(a)-(b). Another observation worth mentioning from figures \ref{core_omega}(b) and \ref{velocity} is that while for $\displaystyle R_{imp}>0.90$, the centers of the stopping vortices starts appearing near the wedge-tip, for $\displaystyle R_{imp}<0.90$, they appear much below it.

For the benefit of the readers, we also present an accompanying video start-stop.avi, which demonstrates the evolution of the starting and stopping vortex for $t_{dec} = 0.002$ side by side with that of $t_{dec} = 0.25$. While for the former, a strong stopping vortex is observed, the latter reveals no stopping vortex. Besides, in figure , we demonstrate the behaviour of the vortices in the long run when the stopping vortex does not form or extremely weak by plotting the streaklines for  $t_{dec} = 0.20$ and $t_{dec} = 0.30$ at time $t=1.10$.

\begin{figure}
	\begin{center}
	\includegraphics[width=0.445\textwidth]{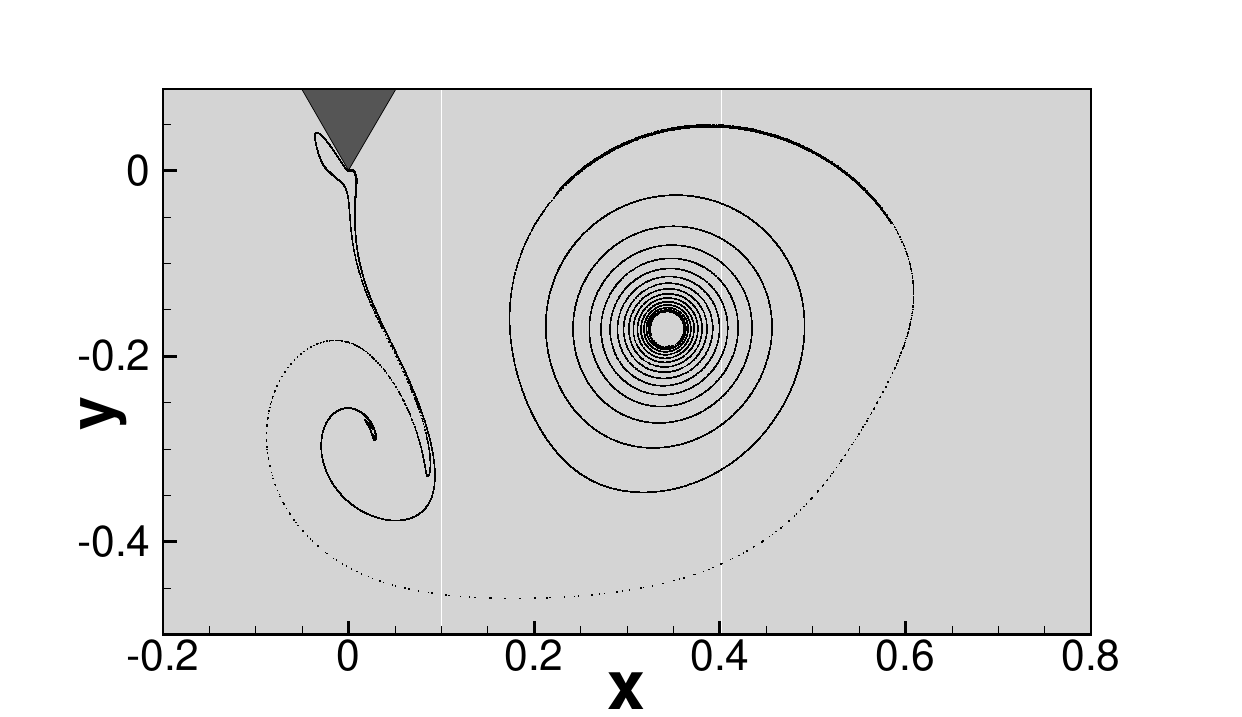}
	{(a)}
	\includegraphics[width=0.445\textwidth]{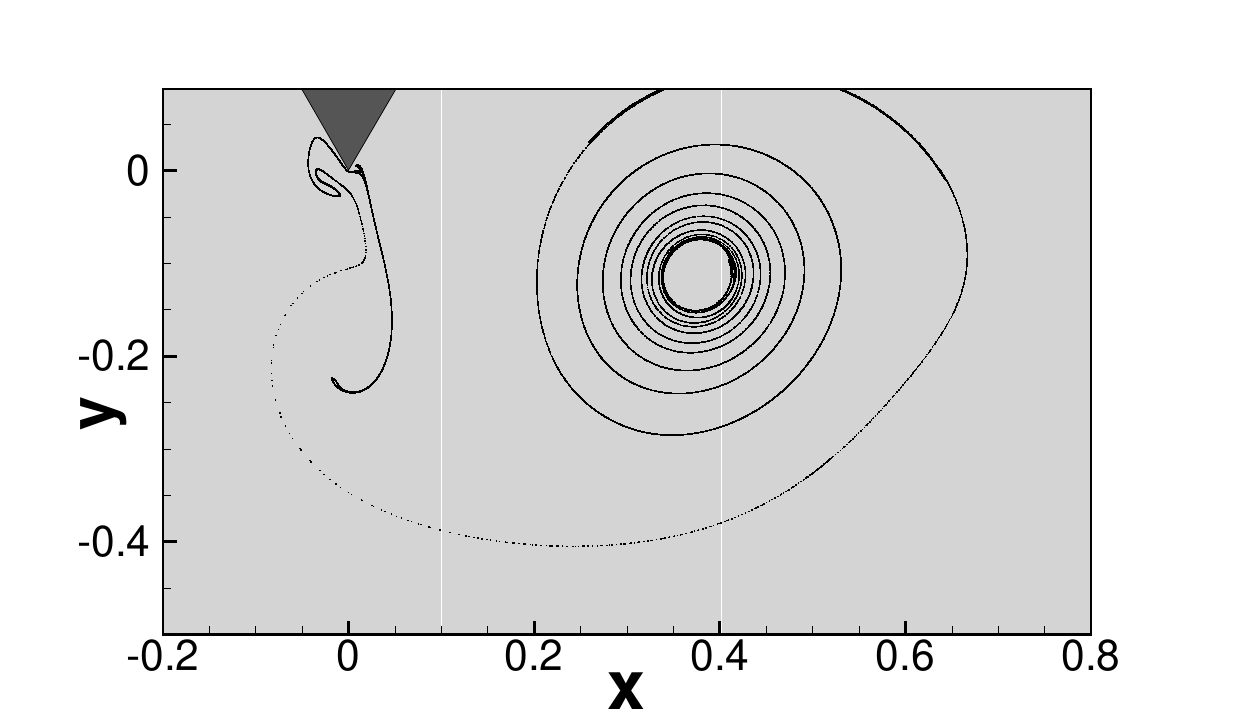}
	{(b)}
	\caption{ Streaklines for  (a) $t_{dec} = 0.20$ and (b) $t_{dec} = 0.30$ at time $t=1.10$. While on the left, the figure indicates an extremely weak stopping vortex, on the right no such formation could be seen.}
	\label{weak_vort}
	\end{center}
\end{figure}

\begin{figure}
	\centering\psfig{file=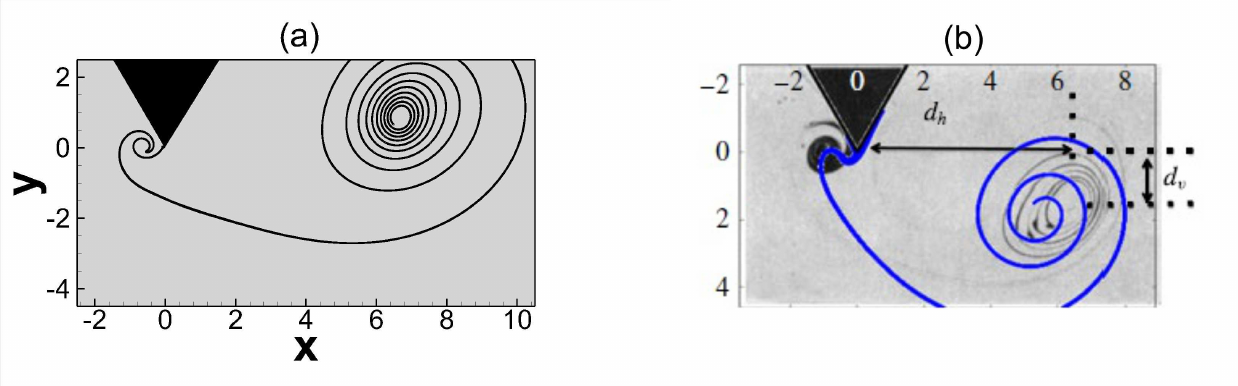,width=\textwidth} 
	\caption{ (a) The streaklines from the current computation at time $t=0.4725$ for $t_{dec}=0.0001$ and   and (b) the streaklines from the computation of  \citep{xu2016} (blue) at the same instant along with the patterns of the fluid dyes in the experimental visualization of \citep{pullin1980} (black) for $Re=1560$. The length in (a) is scaled back to the dimensional one to match those of \citep{pullin1980}.}
	\label{xu_current}
	\end{figure}

It is interesting to note that a clean vortex could not be sighted in perhaps the only endeavour \citep{xu2016} available in the literature to numerically replicate the stopping vortex of \citep{pullin1980} where the flow was stopped at $t=0.4551$. This could be seen from figure \ref{xu_current}(b) where the streaklines from their computation (in blue) was compared with the experimental visualization of \citep{pullin1980} by overlapping the patterns of the fluid dyes with the streaklines. On the other hand, the streaklines from our simulations for $\displaystyle R_{imp}>0.90$ resulted in a clear vortex as evidenced by figure \ref{xu_current}(a) and also the top row of the previous figue, viz., figures \ref{dec_gap}(a)-(d). One can clearly observe a closer proximity of the stopping vortex resulting from our simulation in figure \ref{xu_current}(a) at time $t=0.4725$ to the experimental visualization of \citep{pullin1980}. A larger $t_{dec}$ used  in the simulation of \citep{xu2016} could be a possible reason for the absence of a clear stopping vortex as their streaklines were very similar to the ones in figure \ref{dec_gap} corresponding to $t_{dec}=0.25$ and the accompanying video start-stop.avi. 

 \subsection{The core of the starting vortex}
 \begin{figure}
\begin{tabular}{ccccc}
\hspace{-0.5cm}\epsfig{file=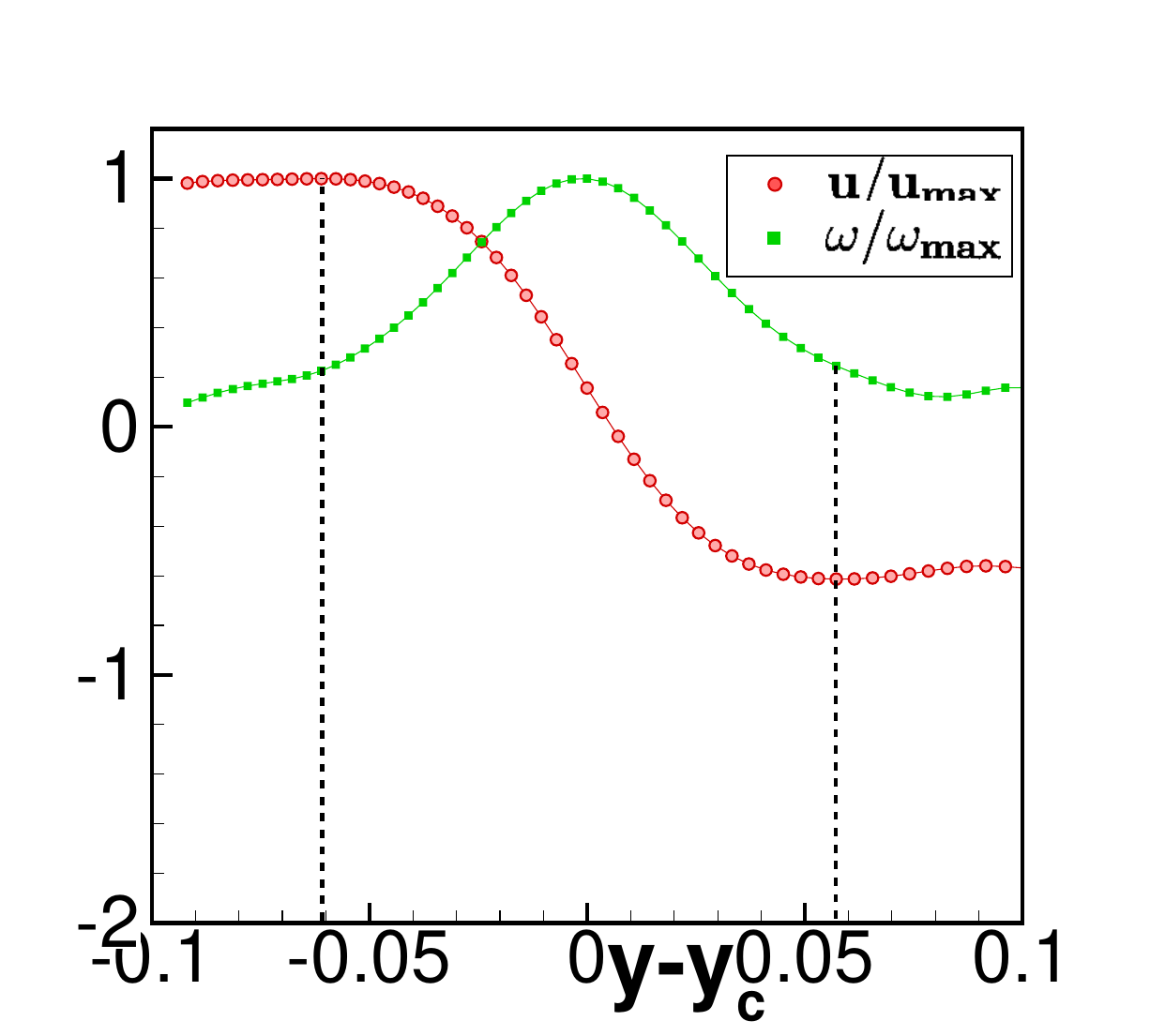,width=0.225\linewidth,clip=}
&
\hspace{-0.5cm}\epsfig{file=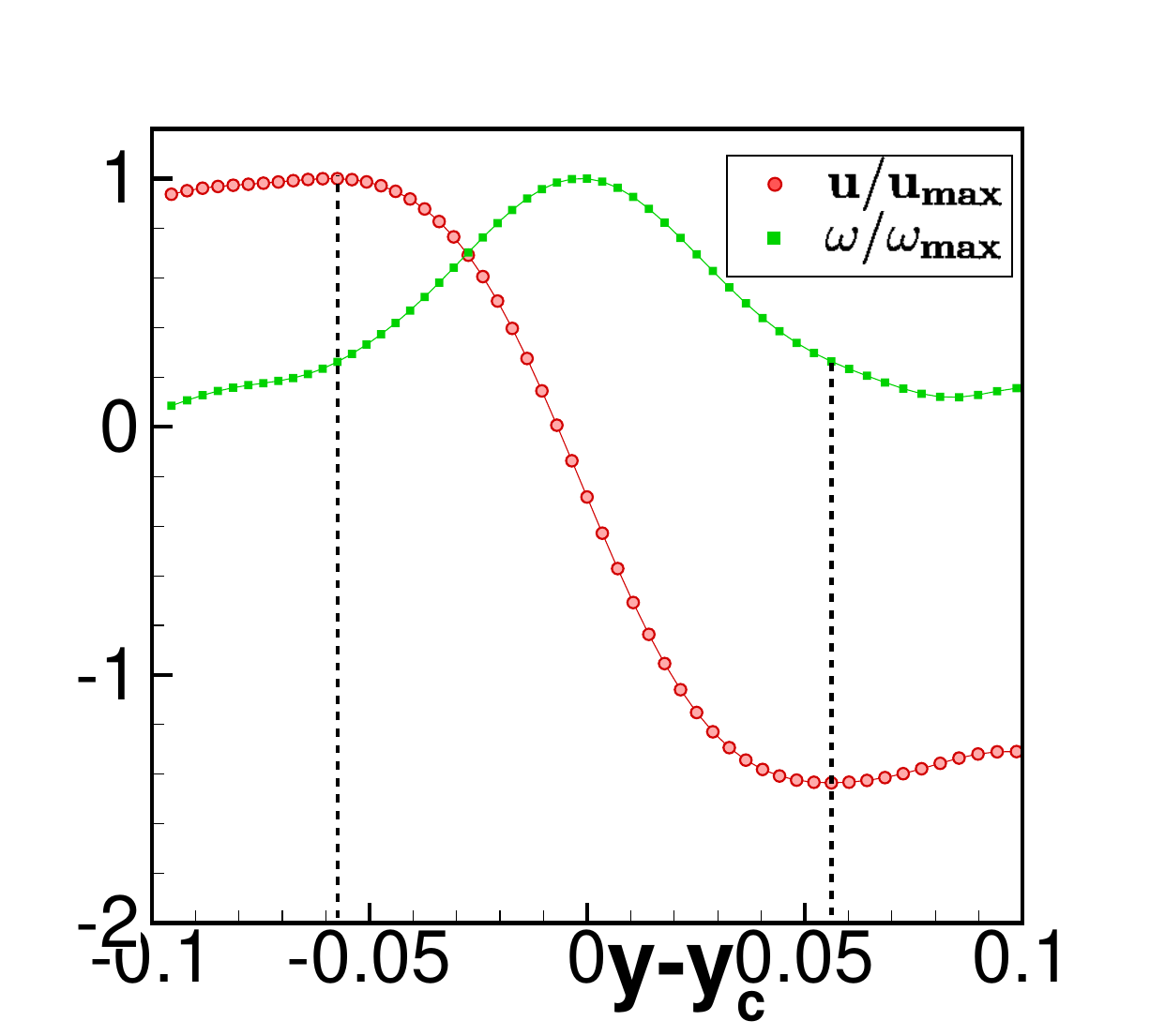,width=0.225\linewidth,clip=}
\&
\hspace{-0.5cm}\epsfig{file=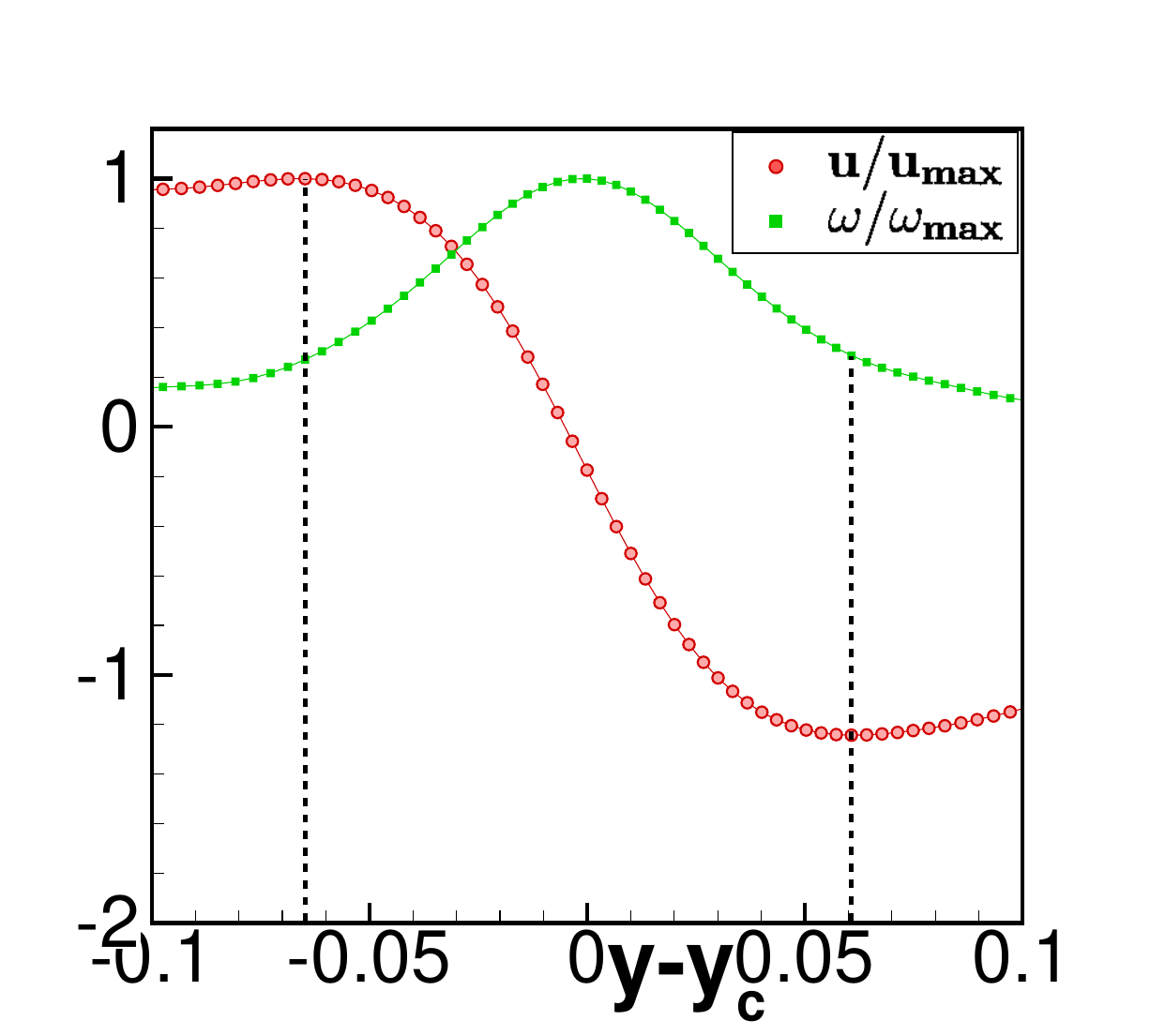,width=0.225\linewidth,clip=}
&
\hspace{-0.5cm}\epsfig{file=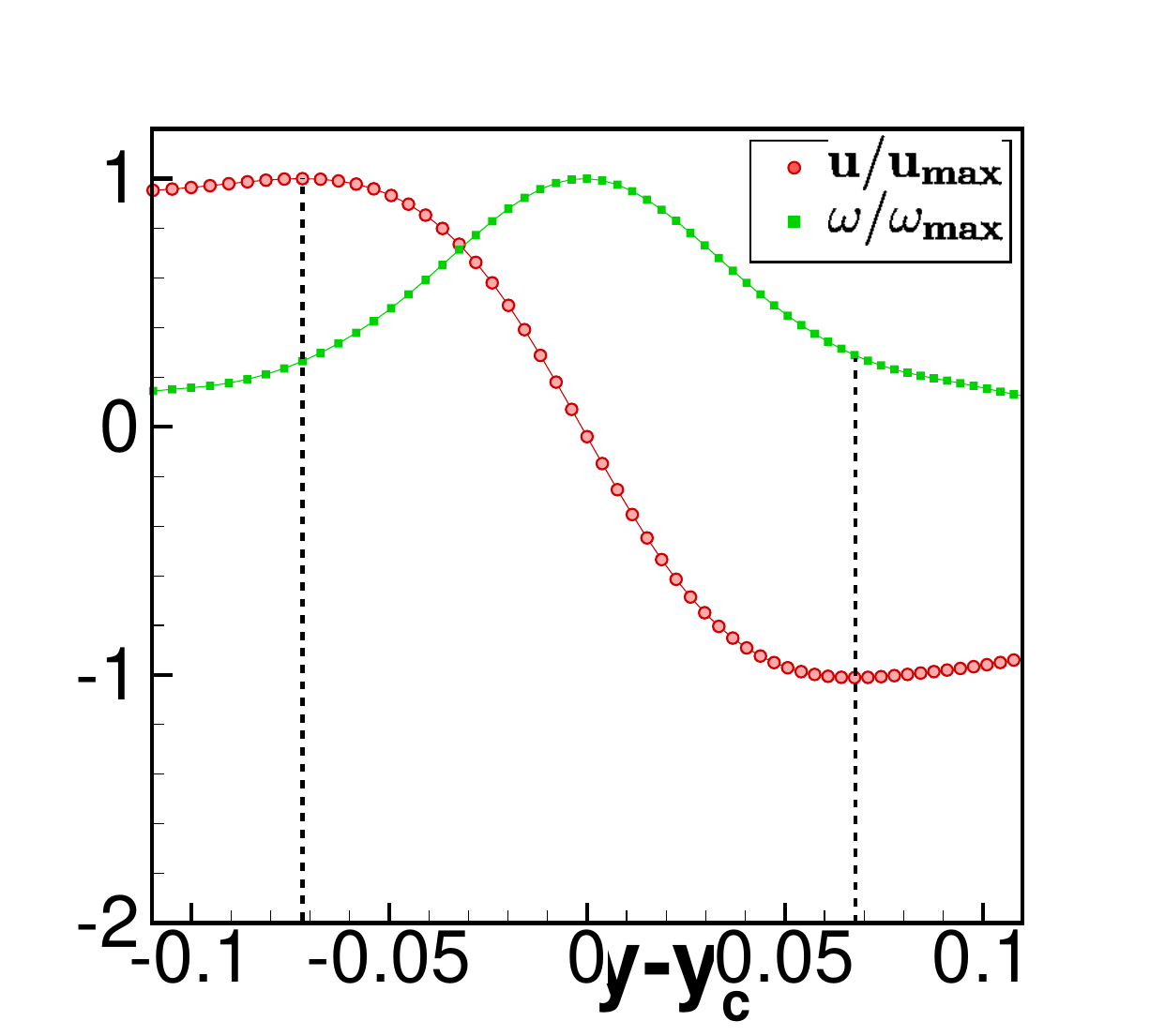,width=0.225\linewidth,clip=}
&
\hspace{-0.5cm}\epsfig{file=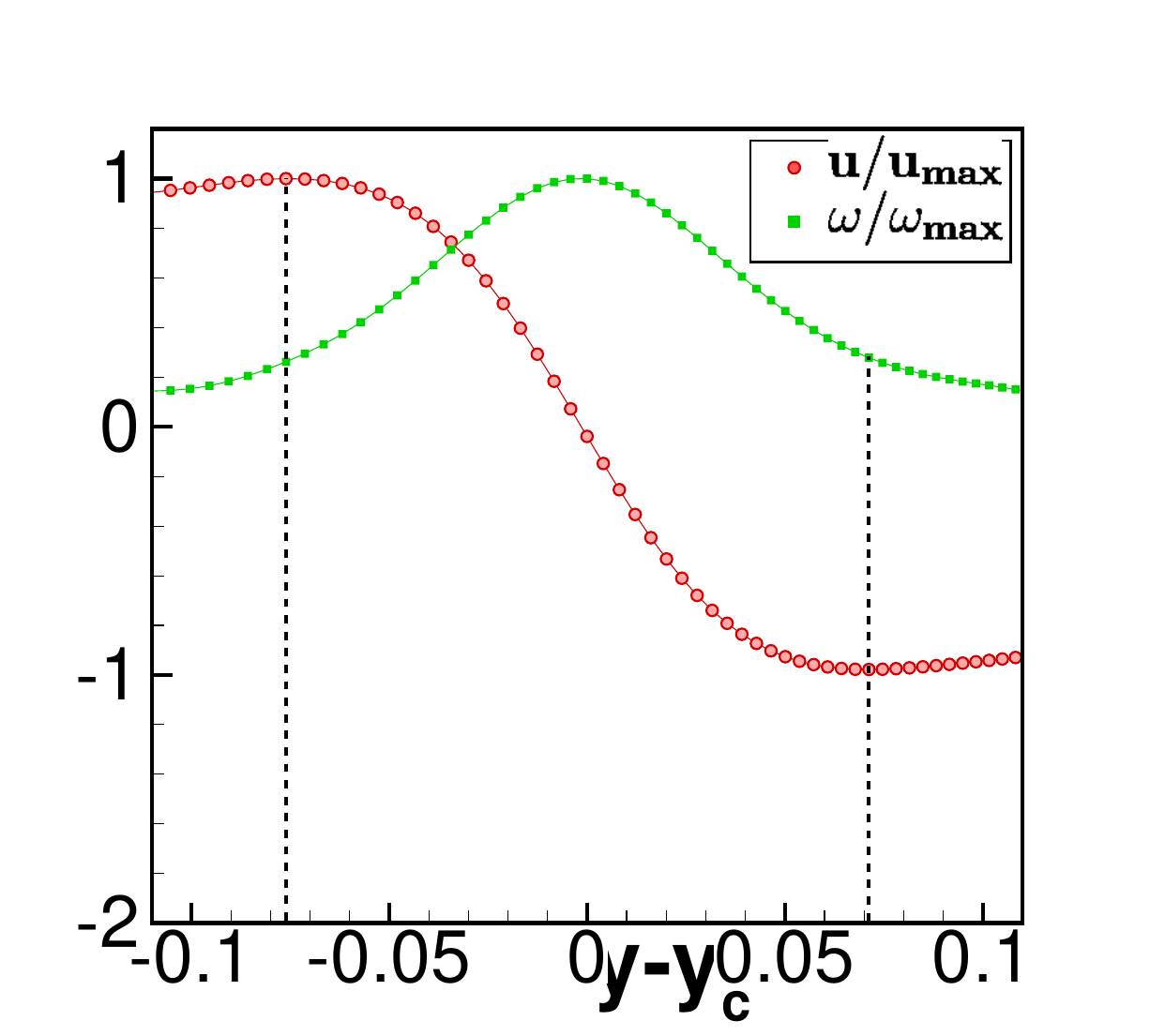,width=0.225\linewidth,clip=}
\end{tabular}
\begin{tabular}{ccccc}
\hspace{-0.5cm}\epsfig{file=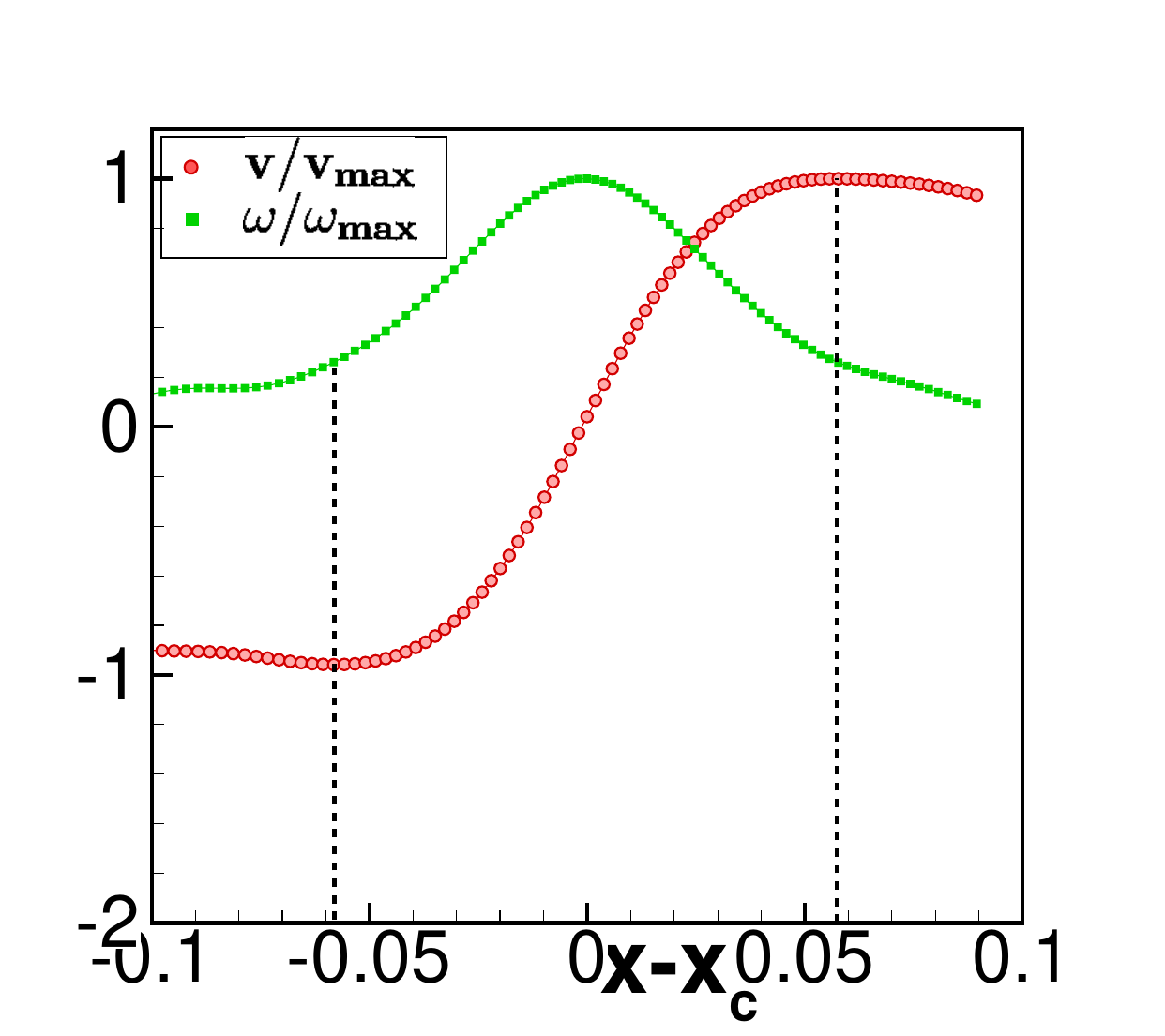,width=0.225\linewidth,clip=}
&
\hspace{-0.5cm}\epsfig{file=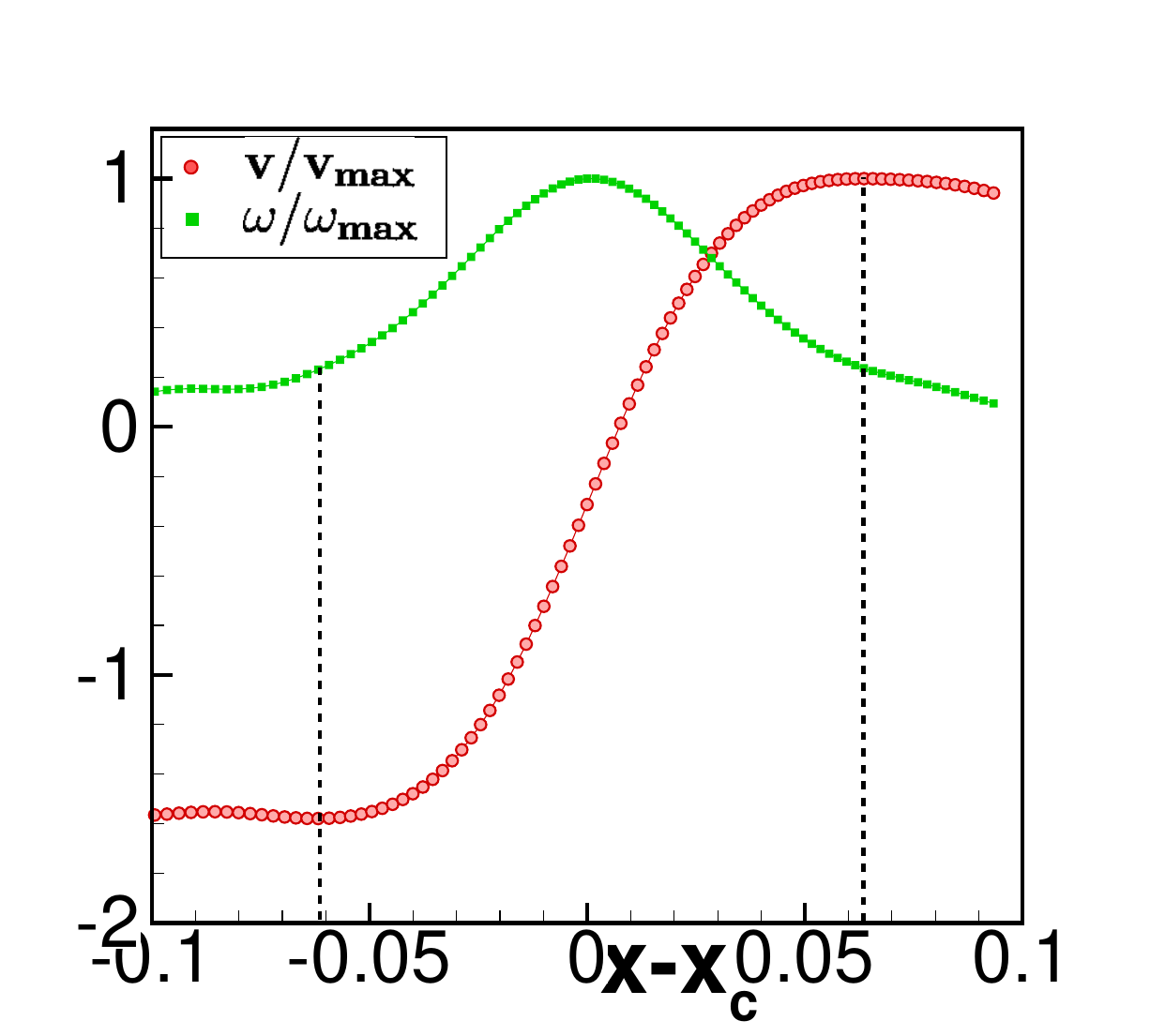,width=0.225\linewidth,clip=}
\&
\hspace{-0.5cm}\epsfig{file=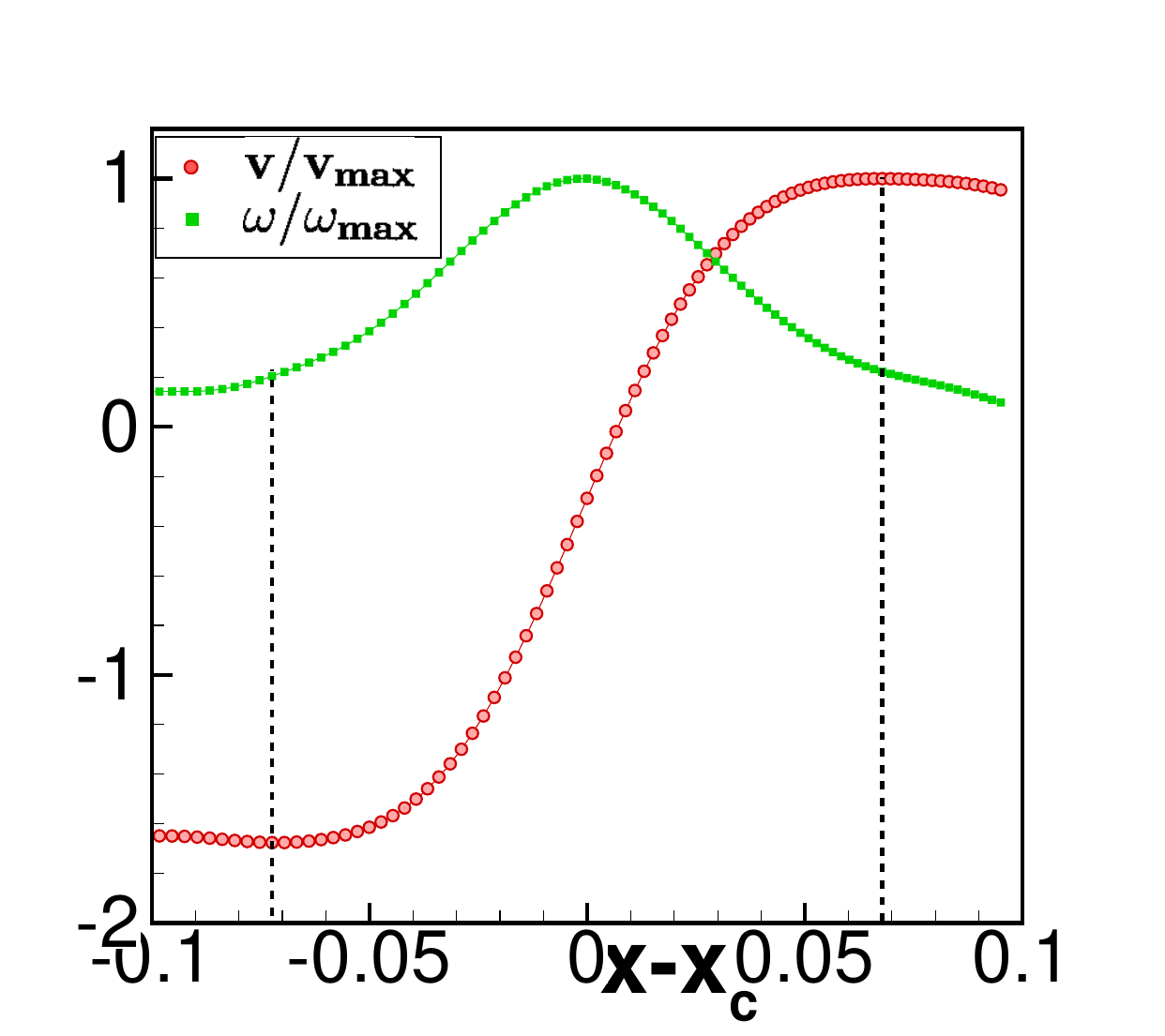,width=0.225\linewidth,clip=}
&
\hspace{-0.5cm}\epsfig{file=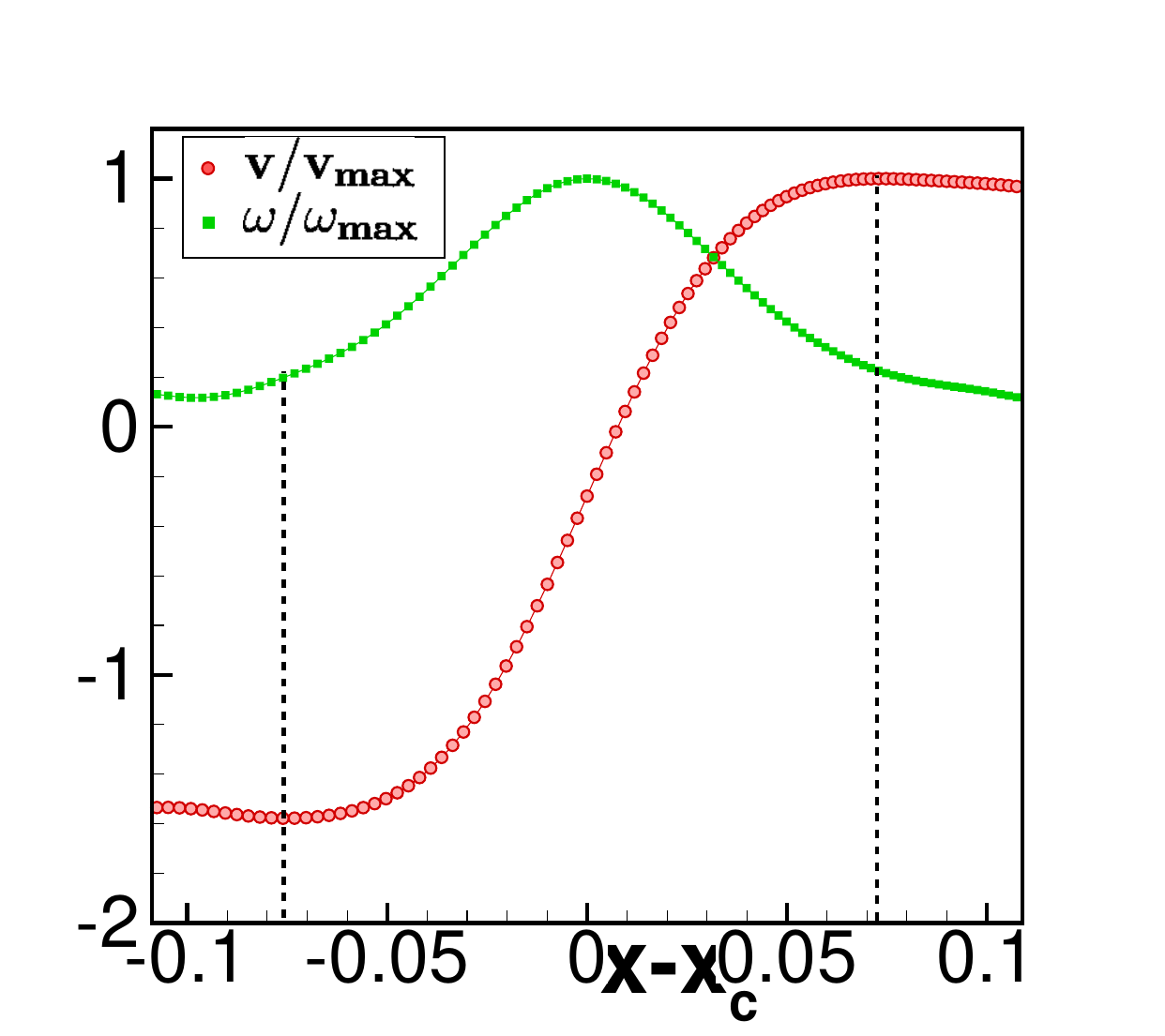,width=0.225\linewidth,clip=}
&
\hspace{-0.5cm}\epsfig{file=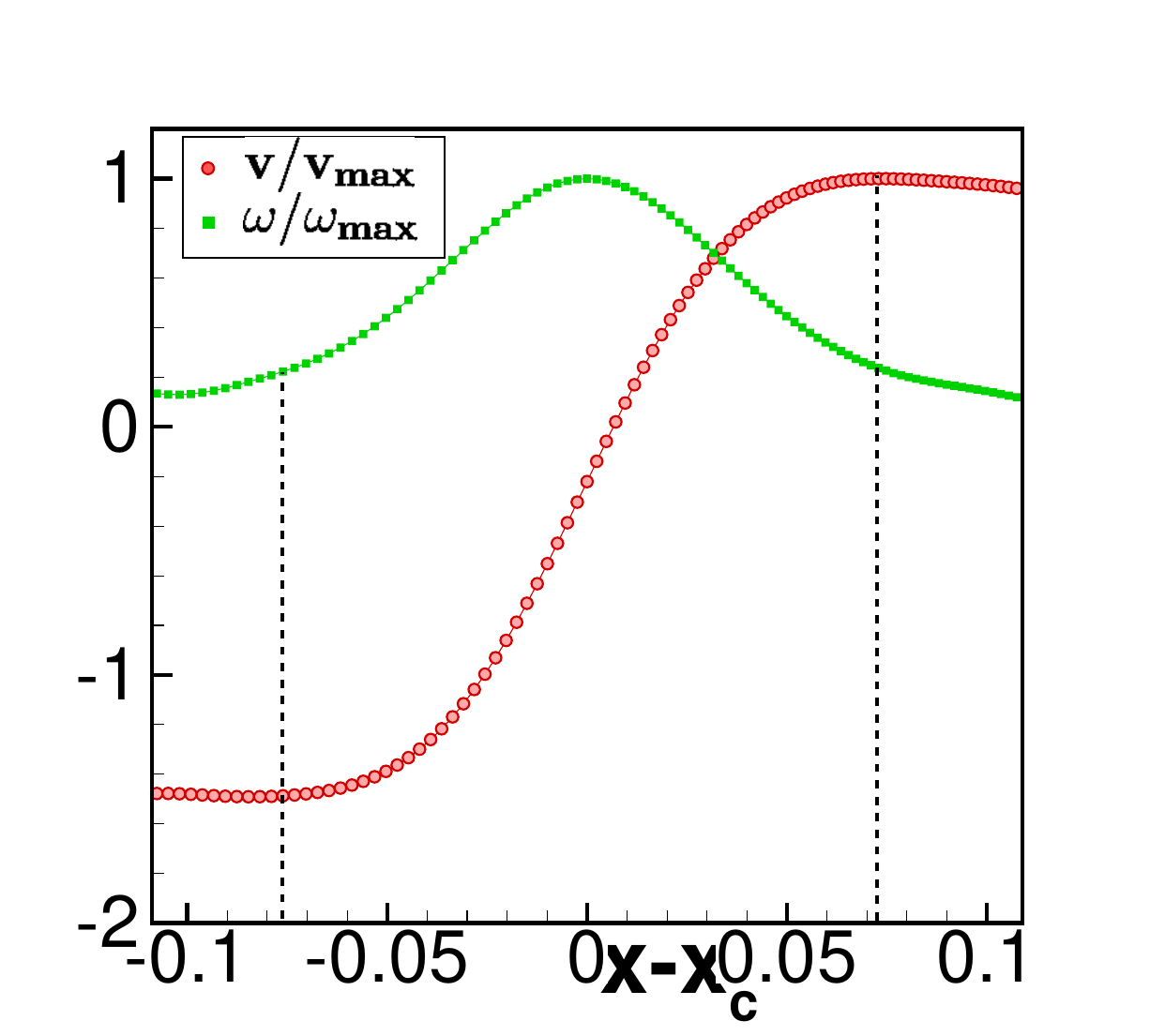,width=0.225\linewidth,clip=}
\end{tabular}
\caption{{The characteristics of the vortex core, from left to right at times $t=0.43$, $0.45$, $0.55$, $0.65$  and $0.70$; $t_{dec}=0.01$: (a) the vorticity and the vertical velocity along the horizontal diameter diameter of the core, and (b) the vorticity and the horizontal velocity along the vertical diameter of the core.}}
\label{core}
\end{figure}
 The extent of the vortex core is indicated by the core diameter, which in turn can be expressed as a function of both the vorticity and velocity distribution. \citep{weigand1997evolution} have defined the  distance between the maxima and minima of the $u$-velocity ($v$-velocity) along the vertical (horizontal) line through the vortex center (see figure \ref{sc_center}) as one of the estimators of the core diameter. The top and bottom rows of figure  \ref{core} respectively exhibit the normalized velocity and vorticity distributions scaled by their respective maximum values along the horizontal and vertical lines through the vortex center at different time stations for $t_{dec}=0.01$. We have shifted to the vortex center to the origin in the plots as depicted by the labels of the horizontal axis. Here, the maxima and minima of the $u$- and $v$-velocities are indicated by drawing vertical lines through their graphs. As one can see from the figures, the vorticity distributions are symmetric about $x-x_c=0$ ($y-y_c=0$) and follow a Gaussian profile. The core diameters estimated along the horizontal directions are very close to the ones obtained along the vertical directions, thus implying an almost circular core region. This observation is consistent with the streaklines presented in figures \ref{dec_gap} and \ref{xu_current}(a). \citep{das2017} observed a drop of $10\%$ of its maximum value at the two extremes of the core diameter. However, our computed results were seen to yield a drop of approximately between $22\%$ to $27\%$ for the same. Also obvious from figure \ref{core} is the occurrence of the vorticity maxima at the points of inflection of the velocity curves. 

 \section{Conclusion}\label{concl}
	This paper is concerned with a numerical study of the starting and stopping flow past a wedge mounted on a wall in a channel for wedge angle $60^{\circ}$ and channel Reynolds number $Re_c=1560$. The simulation is carried out by discretizing the transient Navier-Stokes equations governing the flow in $\psi$-$\omega$ formulation by using a fourth order spatially and second order temporally accurate compact finite difference method on a nonuniform Cartesian grid developed by the author. Our work is inspired by the famous laboratory experiment of starting and stopping vortices by \citep{pullin1980}. In particular, the the dynamics of the stopping vortices have been studied in length as no such details are available from their work. To the best of our knowledge, this is probably the first numerical experiment of successfully mimicking the stopping vortex emanating from exactly the same flow configuration used in laboratory. 

 We validate our numerical simulation by comparing the early evolution of the flow with the experimental visualization of \cite{pullin1980} and performing a grid-independence study. Next, the effect of the time interval $t_{dec}$ through which the inlet velocity of the flow is decelerated to zero is investigated thoroughly in order to analyze the stopping flow. For all the values of $t_{dec}$ considered in the study, the trajectories of both the starting and stopping vortex centers were seen to descend vertically downwards after the incoming flow is stopped. The distance through which this descend occurs, is inversely proportional to the value of $t_{dec}$. The criterion for the development of a clean stopping vortex is provided in terms of the impulse associated with the deceleration. When the impulse corresponding to the deceleration is less than $10\%$ of the total impulse imparted by the incoming flow, a stopping vortex is formed. The strength of the stopping vortex is seen to depend upon the rapidity of deceleration. Its strength was seen to drastically reduced when the impulse ratio falls below the threshold value $0.9$.
 We also provide the characteristics of the main starting vortex after the incoming flow being brought to a complete halt. While the growth of the size of this vortex is almost independent of $t_{dec}$, a smaller value of it was seen to bring down growth rate of the circulation. The vorticity distribution inside the core of this vortex is almost Gaussian and a drop of around $25\%$ of the maximum vorticity was observed inside the core.

 \vspace{.2cm}
	\noindent {\bf Declaration of interest:} The author reports no conflict of interest.
	
	\bibliographystyle{jfm}
	\bibliography{references}
\end{document}